\documentclass[aps,prl,twocolumn,superscriptaddress]{revtex4}
\usepackage{amsfonts}
\usepackage{amsmath}
\usepackage{amssymb}
\usepackage{graphicx}
\setcitestyle{super}

\begin{document}

\title{ Enhanced diffusion in finite size simulations of a fragile diatomic glass-former.}


\author{Sonia Taamalli}
\affiliation{Laboratoire de Photonique d'Angers EA 4464, 
University of Angers, Physics Department,  2 Bd Lavoisier, 49045 Angers, France.}
\affiliation{University of Monastir, Physics Department, Monastir, Tunisia.}

\author{Julia Hinds}
\affiliation{Laboratoire de Photonique d'Angers EA 4464, 
University of Angers, Physics Department,  2 Bd Lavoisier, 49045 Angers, France.}
\affiliation{Appalachian State University, Physics department, Boone, North Carolina, USA.}

\author{Samuel Migirditch}
\affiliation{Laboratoire de Photonique d'Angers EA 4464, 
University of Angers, Physics Department,  2 Bd Lavoisier, 49045 Angers, France.}
\affiliation{Appalachian State University, Physics department, Boone, North Carolina, USA.}

\author{Victor Teboul}
\email{victor.teboul@univ-angers.fr}
\affiliation{Laboratoire de Photonique d'Angers EA 4464, 
University of Angers, Physics Department,  2 Bd Lavoisier, 49045 Angers, France.}


\begin{abstract}
Using molecular dynamics simulations we investigate the finite size dependence of the dynamical properties of a diatomic supercooled liquid. The simplicity of the molecule permits us to access the microsecond time scale. We find that the relaxation time decreases simultaneously with the strength of cooperative motions when the size of the system  decreases. While the decrease of the cooperative motions is in agreement with previous studies, the decrease of the relaxation time opposes what has been reported to date in monatomic glass-formers and in silica. This result suggests the presence of different competing physical mechanisms in the relaxation process.
For very small box sizes the relaxation times behavior reverses itself and increases strongly when the box size decreases, thus leading to a non-monotonic behavior. This result is in qualitative agreement with defect and facilitation theories.

\end{abstract}

\maketitle


\section{Introduction}

The microscopic origin of the large increase of the relaxation times of supercooled liquids upon decreasing the temperature is still a matter of conjecture\cite{anderson,gt1,gt2,gt4}. Most theories \cite{adam,chandler,facilitation,defect,gt2,RFOT,RFOT2,frust,frust2,MCT} postulate that this dynamical slowing down is related to the increase of a correlation length, possibly due to cooperativity\cite{dh1,dh2,dh3,dh4,dh5}, and that it may have a structural or dynamical origin. 
This picture of cooperative length scales is supported by the increase of the activation energy in fragile glass-formers\cite{fragile1,fragile2} to values commonly larger than a typical chemical bond energy, leading to a non-Arrhenius dependence of the $\alpha$ relaxation time with the temperature.

To better understand the relation between the cooperativity and the viscosity, the simplest method is to modify the cooperativity and study the effect of induced modification on the viscosity.
Because the $\alpha$ relaxation time $\tau_{\alpha}$ and the viscosity $\eta$ are related, (most common assumptions on that relation are $\tau_{\alpha} \sim \eta/T$ or $\tau_{\alpha} \sim \eta$, for a critical review see \cite{shi}),   we chose to focus our attention on the relaxation time $\tau_{\alpha}$ in this work.
In molecular dynamics simulations, the system size introduces a cutoff on any cooperative mechanism and can thus be used to tune the cooperativity\cite{mdgt4,size0,size1,size2,size3,size4,size5,size6,size7,size10,size8}. 
Since the cooperative motions and the viscosity increase simultaneously when the temperature drops, we may expect a decrease of the viscosity when the system size,  and consequently the cooperativity, will decrease. 
Models predicting that kind of  behavior are\cite{size2} the Adam-Gibbs theory\cite{adam}, the frustration limited domain theory\cite{frust,frust2}, the random first order transition theory (RFOT)\cite{RFOT,RFOT2,gt2}, and the facilitation theory, while the Mode Coupling Theory predicts the opposite effect\cite{MCT}. 
The facilitation theory\cite{facilitation} and more generally  defect theories\cite{defect} predict also an abrupt slowing down when the size of the system will be small enough for no defect (no excitation in the facilitation theory) to be temporarily present in the system.
A review of the different theories predictions on size effects can be found in ref.\cite{size2}. 
Note that a  different picture on cooperative mechanisms predicts an increase of the viscosity when the system size decrease.
When the temperature drops it becomes increasingly harder to find relaxation pathways with low enough energy barriers, leading to an increase of the relaxation time. Finding those pathways requires considering cooperative rearrangements involving larger and larger numbers of particles. Truncating the system size thus removes the possibility of finding such large-scale cooperative events, leading to an increase of the relaxation time in that picture.

Molecular dynamics (M.D.) simulations\cite{md1,md2,md3} reproduce the unexplained slowing down at the approach of the glass transition temperature, while having the unique property to give access to the position and motion of each atom at any time during the virtual experiment.
MD simulation is thus an invaluable tool to study the glass-transition problem\cite{mdgt2,mdgt3,mdgt5,mdgt6,mdgt7} and more generally condensed matter phenomena\cite{mdcm1,mdcm2,mdcm3,mdcm4}. 
In this work we study finite size effects using molecular dynamics simulations of a simple molecular liquid. We use finite size simulations\cite{mdgt4,size0,size1,size2,size3,size4,size5,size6,size7,size10,size8} instead of  confinement\cite{pore1,pore2,pore3,pore4,pore5,pore6,pore7,pore8,pore9,pore10,pore11,pore12} because finite size simulations  have the advantage over confinement to  cut off the cooperativity  without introducing any confining wall nor modifying the symmetry or the dimensionality of the system, as long as periodic boundary conditions are used.
The simplicity of the molecule permits us to access large time scales with aging times larger than the micro-second.

Previous experiments and simulations using confinement found, depending on the conditions, mostly an increase, but sometimes a decrease of the viscosity with the system size\cite{pore1,pore2,pore3,pore4,pore5,pore6,pore7,pore8,pore9,pore10,pore11,pore12}.
However previous finite systems simulations \cite{mdgt4,size0,size1,size2,size3,size4,size5,size6,size7,size10,size8} found to our knowledge always an increase of the viscosity together with a decrease of the cooperativity when the system size decreases.
In this work we observe for the first time, a decrease of the relaxation times associated with a decrease of the cooperativity when the system size decreases. When decreasing the system size further we observe the abrupt slowing down predicted by facilitation and defects theories. We also observe the intermittent disappearance of the excitations in the system sizes corresponding to that strong slowing down.


\section{Calculations}

We model the molecules\cite{ariane} as constituted of two atoms ($i=1, 2$) that we rigidly bond fixing the interatomic distance to $d=2.28 $\AA$ $.
 Each atom of our linear molecule has a mass m=50 g/$N_{a}$ where $N_{a}$ is the Avogadro number.
Atoms of the set of linear molecules constituting our liquid interact with the following Lennard-Jones potentials: 
\begin{equation}
V_{ij}=4\epsilon_{ij}((\sigma_{ij}/r)^{12} -(\sigma_{ij}/r)^{6}) 
\end{equation}

 with the parameters: $\epsilon_{11}= \epsilon_{12}=0.5$ kJ/mol, $\epsilon_{22}= 0.4$ kJ/mol, $\sigma_{11}= \sigma_{12}=4.56$\AA\ and $\sigma_{22}=4.33$\AA$ $.
   
Our molecule is thus $6.7$\AA$ $ long and $4.5$\AA$ $ wide.
With these parameters the liquid does not crystallize when supercooled even during long simulation runs\cite{ariane}. This model has been described and studied in detail previously\cite{ariane} and was found to display the typical behaviors of fragile supercooled liquids.
To increase the efficiency of our simulations and be able to decrease significantly the simulation box size,  we chose a small cutoff value for the Lennard Jones potentials $r_{c}=1.55$ $ \sigma_{11} = 7.07$ \AA. 
The potential function is then shifted as usual so that its value at the cutoff is zero. This method insures that no energy fluctuations are induced by the cutoff.
We found very small differences between simulations with larger cutoff values (2.5$\sigma_{11}$ and 3$\sigma_{11}$) and our small cutoff (1.55$\sigma_{11}$), in the conditions T=120K and N=500 molecules.
Note that more sophisticated methods exist that suppress the force discontinuity at the cutoff\cite{cutoff} but modify the shape of the potential.
 As they are modeled with Lennard-Jones atoms, the potentials are quite versatile.
Due to that property, a shift in the parameters $\epsilon$ will shift all the temperatures by the same amount, including the glass-transition temperature and the melting temperature of the material. We use periodic boundary conditions.
The density is set constant in our calculations at $\rho=1.615 g/cm^{3}$. When rescaled, or in dimensionless units, that density value is larger than the density of the original model\cite{ariane}, and thus leads to a more viscous medium. For large system sizes, the relaxation time $\tau_{\alpha}$ follows the Mode Coupling Theory behavior $\tau_{\alpha}=\tau_{0}$ $(T-T_{c})^{-\gamma}$ with a critical temperature  $T_{c}=106 K$. 
We use the Gear algorithm with the quaternion method\cite{md1} to solve the equations of motions with a time step $\Delta t=10^{-15} s$. 
The temperature is controlled using a Berendsen thermostat\cite{berendsen}. 
We give more details in the appendix.  

\section{Results and discussion}

\subsection{Dynamical and structural evolution with the system size:\\}

We display in Figure 1 the incoherent scattering function  $F_{S}(Q,t)$ which represents the autocorrelation of the density fluctuations at the wave vector Q.
This function gives information on the structural relaxation of the material.
We define $F_{S}(Q,t)$ by the relation:
\begin{equation}
\displaystyle{F_{S}(Q,t)={1\over N N_{t_{0}}} Re( \sum_{i,t_{0}} e^{i{\bf Q.(r_{i}(t+t_{0})-r_{i}(t_{0}))}}  )          }\label{e1}
\end{equation}
Here we choose Q as the wave vector corresponding to the maximum of the structure factor $S(Q)$.
$F_{S}(Q,t)$ then allows us to calculate the $\alpha$ relaxation time $\tau_{\alpha}$ from the equation: 

\begin{equation}
\displaystyle{F_{S}(Q,\tau_{\alpha})=e^{-1}}            
\end{equation}

Figures 1a and 1b show the appearance of size effects, since the function $F_{S}(Q,t)$ and its relaxation time depend on the system size, and $\tau_{\alpha}$ increases with the system size in the Figure. 
The function $F_{S}(Q,t)$ in Figure 1a also displays the three time regimes, that are characteristics of supercooled liquids. The molecular motions are ballistic at small times because the molecules have not encountered the boundaries of the cages created by their neighbors. Then, on the plateau time scale, the molecules are trapped inside the cages. Eventually, on the $\alpha$ relaxation time scale, the molecules escape the cages.
Figure 1a shows that size effects are absent on both the ballistic and the plateau regime in our system.
Size effects occur at the plateau ending, i.e. when molecules begin to escape the cages, which suggests that size effects mainly affect the probability to escape the cages, not the microscopic motions of the molecules, or the average size of the cages since the height of the plateau is not modified.
 Since cooperative motions also appear on the same time scale (at the end of the plateau regime) in glass-formers, this result agrees with the hypothesis of a relation between size effects and cooperative motions.

\includegraphics[scale=0.33]{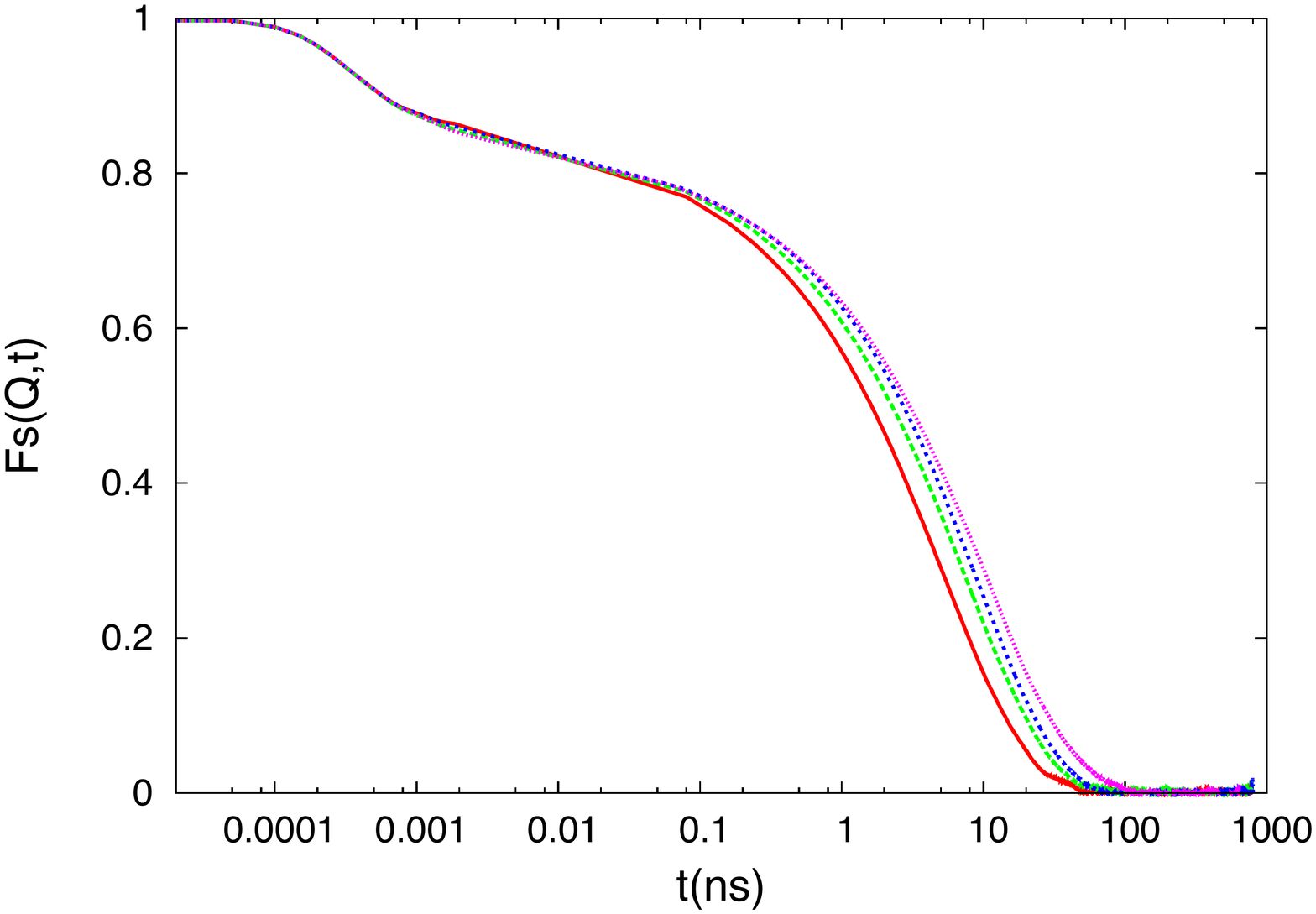}

{\em \footnotesize  FIG.1 (a). (color online)  Self incoherent scattering function $F_{S}(Q,t)$ for the molecules centers of masses using various simulation boxes sizes. We have used the wave number $Q=Q_{0}=1.54$\AA$^{-1}$ that corresponds to the location of the maximum of the structure factor.
The temperature is T=120K. From the left to the right around t=5 ns : N=100 molecules (continuous red line), then N=200, 400, 600 and 800 molecules.
The $\alpha$ relaxation time $\tau_{\alpha}$ increases with the system size.\\}

The $\alpha$ relaxation time $\tau_{\alpha}$  decreases as the system size decreases.
When we decrease the simulation box size, we cut off the cooperative motions in our liquid. Thus, if we interpret the variations of the viscosity as arising from the modification of the cooperativity, for that simple molecular liquid the viscosity increases when the cooperative motions increase, which is similar to what we observe when we decrease the temperature of supercooled liquids.
This result agrees with theories of the glass-transition that are based on activated cooperative mechanisms. 
A well known example is the Adam-Gibbs theory, for which the size of the postulated cooperatively rearranging regions (CRR) decreases when the system size decreases, leading to a decrease of the viscosity.
However previous simulations\cite{mdgt4,size0,size1,size2,size3,size4,size5,size6,size7,size10} on different molecular and atomic liquids have shown an increase of the relaxation times when the system size decreases, results that better agree with the mode coupling theory (MCT).
These two opposite behaviors suggest that different relaxation mechanisms could be present in different supercooled liquids.
 Simulations of slightly different models suggest that these different behaviors are triggered by the shape of the molecule.

\includegraphics[scale=0.31]{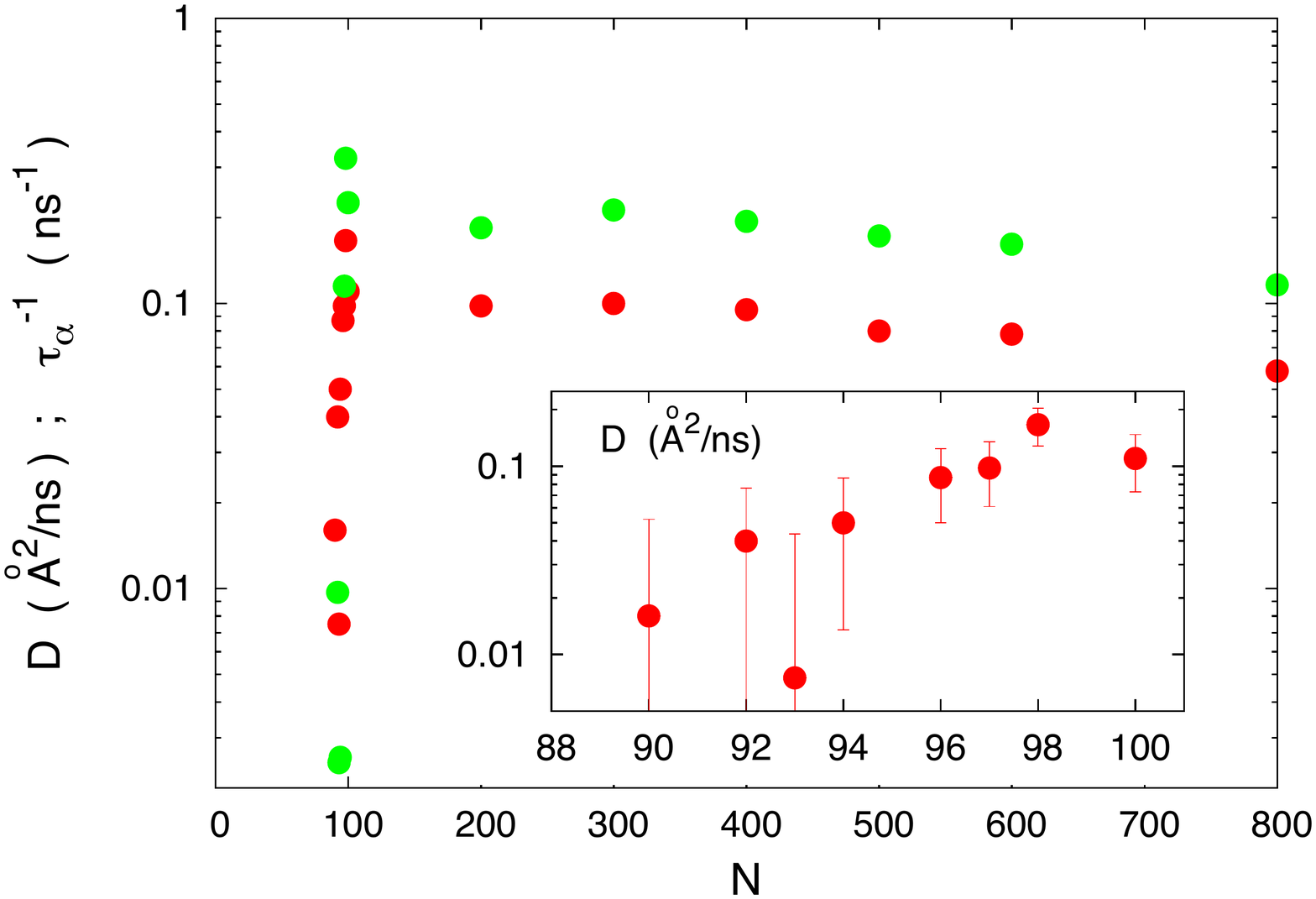}

{\em \footnotesize  FIG. 1 (b). (color online) Diffusion coefficient ($D$: red circles) and inverse of the $\alpha$ relaxation time ($\tau_{\alpha} ^{-1}$: green circles) versus the simulation box size quantified by the number of molecules N. The temperature is T=120 K.  Inset: Detail of the diffusion coefficient $D$ evolution as a function of $N$ for small system sizes, showing the rapid decrease of the diffusion coefficient for $N<98$. This decrease is expected by the facilitation theory for small system sizes, however other causes are possible as for example an undetected partial crystallization of the liquid.}

We observe in Figure 1b an abrupt arrest of the $\alpha$ relaxation when we decrease the system size even further, i.e. below N=97 molecules.
For these small system sizes, when we decrease the temperature to 120K,  the relaxation is first roughly similar to the relaxations of slightly larger boxes (N=100), and then undergoes rapid aging towards slow relaxation. 
This slowing down implies a non-monotonic behavior of $\tau_{\alpha}$, as the relaxation time first decreases with the system size and then strongly increases, supporting the presence of different relaxation mechanisms. We note that the facilitation theory predicts that for small enough systems when defects called excitations will not be present inside the small simulation box the dynamics will stop. Consequently, the behavior we observe, agrees with the facilitation theory prediction for small systems.
However the slowing down for small $N$ is so rapid that one may think of  simpler causes, as for example a partial crystallization of the liquid.
Following that idea, we have searched for partial crystallization that could explain the abrupt slowing down of the dynamics. We didn't find any sign of crystallisation as a modification of the radial distribution function nor large fluctuations in statistical quantities that usually appear during crystallization. However 
similar results on different systems would be needed to insure that this behavior is actually induced by the decrease of the excitation concentrations for small systems. 

\includegraphics[scale=0.33]{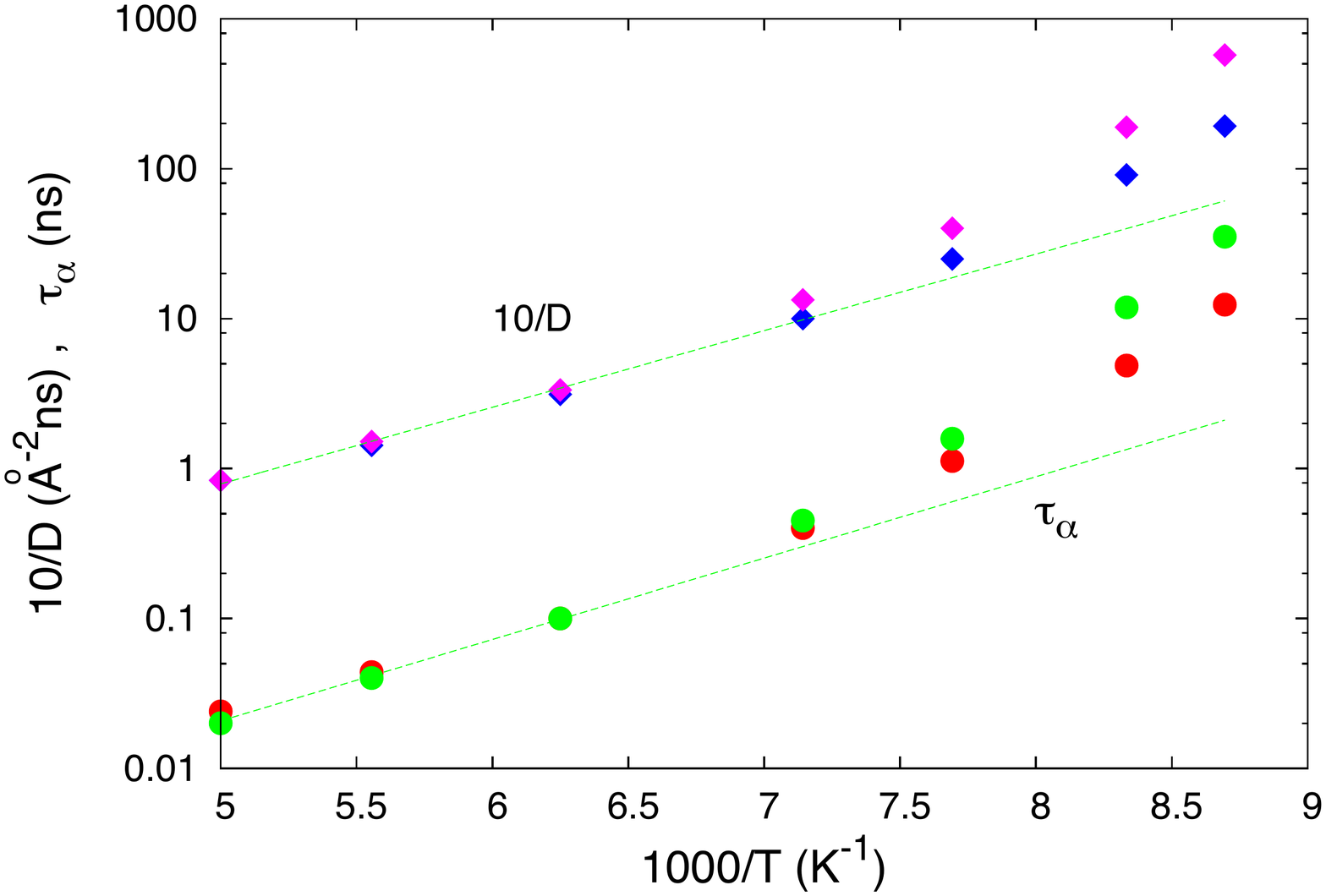}

{\em \footnotesize  FIG. 2. (color online) Diffusion coefficient ($D$: diamonds) and $\alpha$ relaxation time ($\tau_{\alpha}$: circles) versus temperature for different simulation box sizes. The lines correspond to an Arrhenius behavior. The upper curves (purples diamonds and green circles) correspond to the larger boxes i.e. N=800 molecules, while N=100 molecules for the two other curves (blue diamonds and red circles). The larger boxes display the largest relaxation times and smallest diffusion.}

 The non-Arrhenius dependence of the relaxation times and of the diffusion coefficient with temperature is another behavior that most theories associate with the cooperativity in supercooled liquids.  For various values of the box sizes, Figure 2 shows the temperature evolution of the inverse of the diffusion coefficient $1/D$ and of the alpha relaxation time $\tau_{\alpha}$.
We see on the Figure that while the evolution is super-Arrhenius for the larger boxes $\tau_{\alpha}=\tau_{0}$ $e^{E_{a}(T)/k_{B}T}$ (i.e. evolves faster than a pure exponential, or equivalently $E_{a}(T)$ increases with $1/T$), as the size of the box decreases the result tends to an Arrhenius law (i.e. $E_{a}\approx$ constant).
As most theories expect the super-Arrhenius behavior of the diffusion in glass-formers to be due to the increase of a correlation length, the observed decrease of the non-Arrhenius behavior, here associated with a decrease of the cooperative motions, agrees with that picture.
When decreasing the size of the box, the correlation length scale cannot expand to distances further than the box size, thus leading to a more Arrhenius behavior.

\includegraphics[scale=0.33]{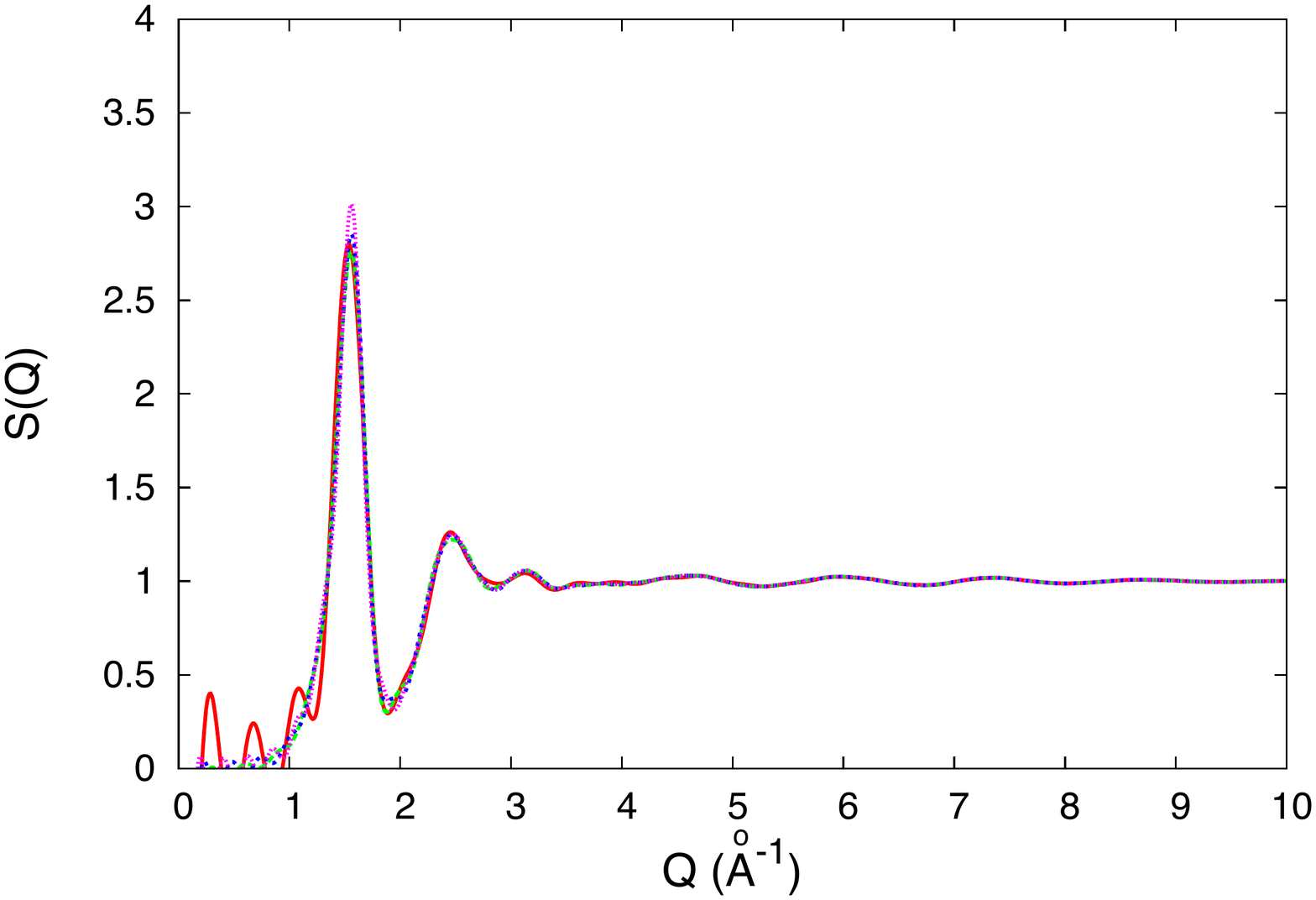}

{\em \footnotesize  FIG.3. (color online)  Structure factor $S(Q)$  for various simulation boxes sizes from $N=100$ (red continuous line) to $N=800$ (purple dotted line). The maximum of the first peak is located at $Q_{0}=1.54$\AA$^{-1}$.
The temperature is T=120K.\\}

In opposition with the dynamical functions studied above, the structure factor $S(Q)$ stays constant when we modify the system size, as shown in Figure 3.
$S(Q)$ reaches its maximum at the wave vector $Q_{0}=1.54$ \AA$^{-1}$, that corresponds to a structural length scale $\displaystyle{\delta_{0}= {2\pi \over Q_{0}}=4.1}$\AA.
We thus do not find any significant variation of that structural length $\delta_{0}$ with the box size, or with the temperature.

\subsection{ Cooperative motions dependence on the system size:\\}

In agreement with previous studies\cite{mdgt4,size0,size1,size2,size3,size4,size5,size6,size7,size10,size8} our results suggest that the strength of the cooperativity decreases when the size of the system decreases.
This decrease of the cooperativity arises from the cutting off of long range cooperative motions, which is due to the decrease of the system length scale.
In the following subsections, we use various correlation functions to directly study the extent of cooperativity inside the liquid, thus showing it actually decreases with the size of the box.

\subsubsection{Dynamic susceptibility:}
For measuring the strength of the cooperative motions, the most convenient function is the dynamic susceptibility $\chi_{4}$ defined as\cite{dh1}:

\begin{equation}
\chi _{4}(a,t)=\frac{\beta V}{N^{2}}\left( \left\langle
C_{a}(t)^{2}\right\rangle -\left\langle C_{a}(t)\right\rangle ^{2}\right) 
\label{e2}
\end{equation}
with 
\begin{equation}
C_{a}(t)={\sum_{i=1}^{N}{w_{a}}}\left( \left\vert {{{\mathbf{r}}}}_{i}(t)-{{{\mathbf{r}}}}_{i}(0)\right\vert \right) .  \label{e3}
\end{equation}

In these equations, $V$ denotes the volume of the simulation box, $N$ denotes the number of molecules in the box, and $\beta =(k_{B}T)^{-1}$. Also, the symbol $w_{a}$ stands for a discrete mobility window function, $w_{a}(r)$, taking the values $w_{a}(r)=1$ for $r<a$ and zero otherwise. We use the value $a=1.5$\AA $ $ in this work, which maximizes $\chi_{4}$ in our liquid at the density of the study.

\vskip1cm

\includegraphics[scale=0.32]{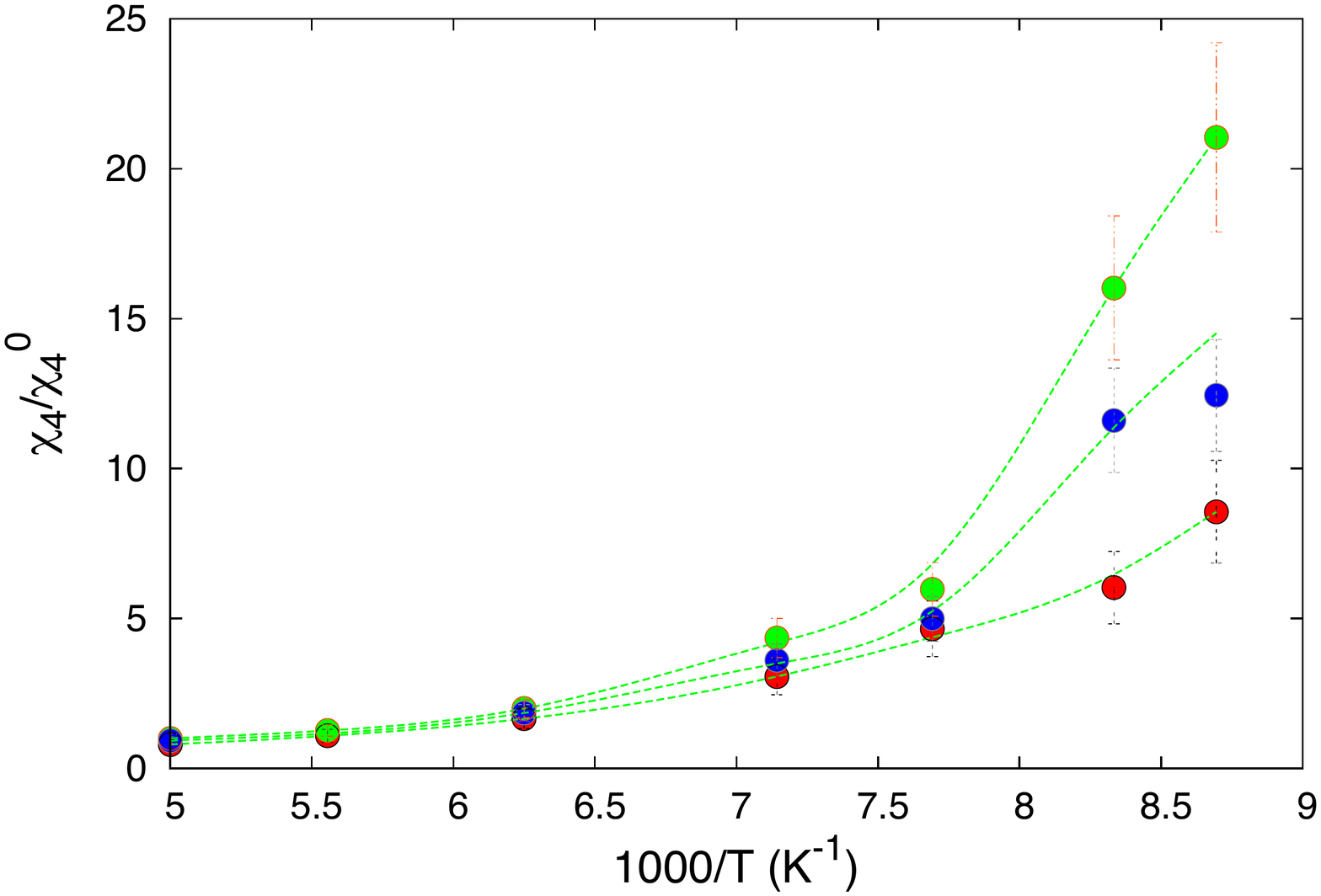}

{\em \footnotesize  FIG.4a. (color online) Maximum value of the 4-point dynamic susceptibility $\chi_{4}$, for different temperatures and system sizes. From top to bottom: N=800 (green circles), N=200 (blue circles), and N=100 molecules (red circles). The lines are guides to the eyes intended to clarify the Figure.\\}

\includegraphics[scale=0.32]{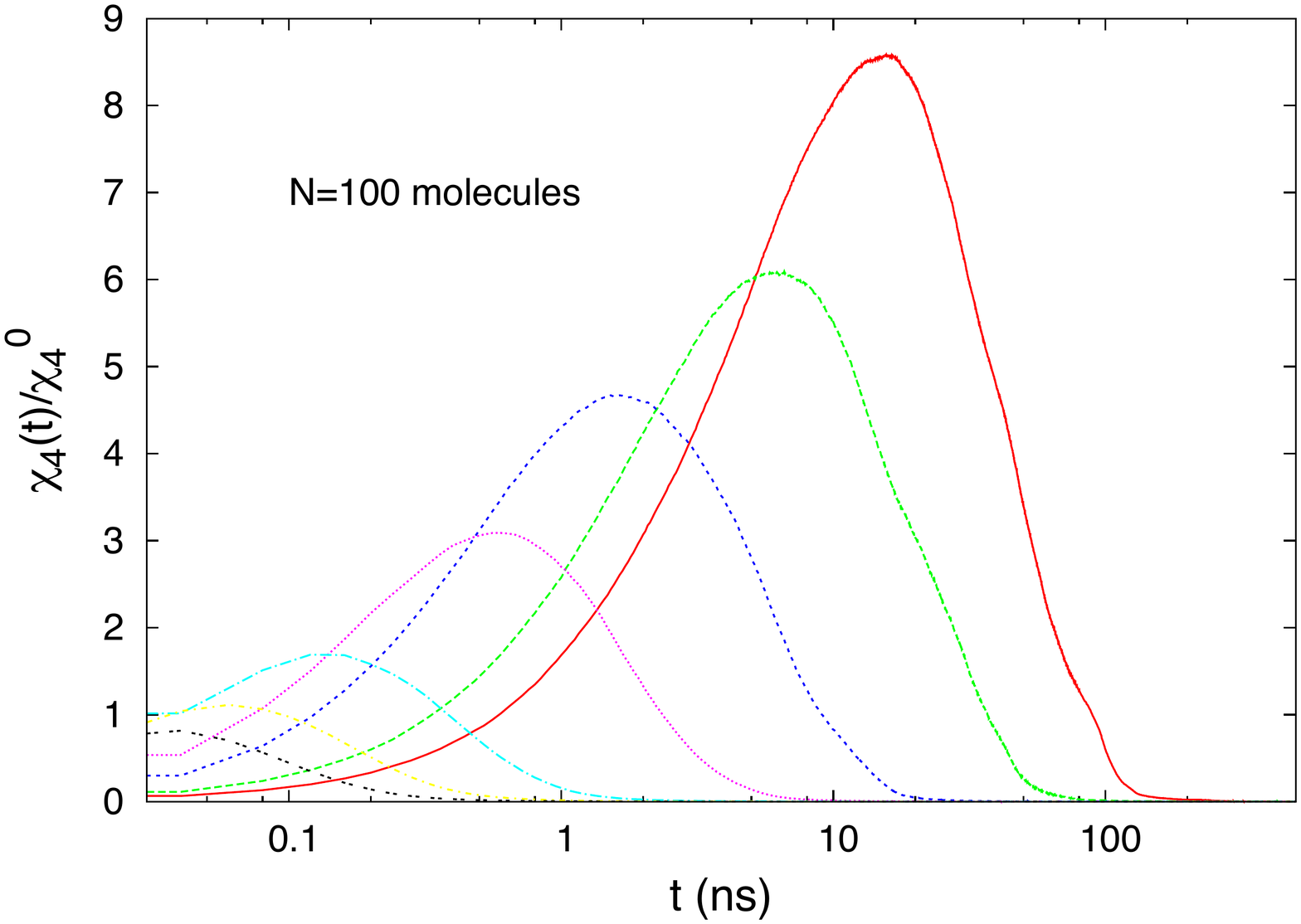}

{\em \footnotesize  FIG.4b. (color online) 4-point dynamic susceptibility $\chi_{4}$, for different temperatures. N=100 molecules. The temperature increases from right to  left: T=115K (red curve), 120K (green curve), 130K (blue curve), 140K (purple curve), 160K (light blue curve), 180K (yellow curve) and 200K (black curve).\\}

\includegraphics[scale=0.32]{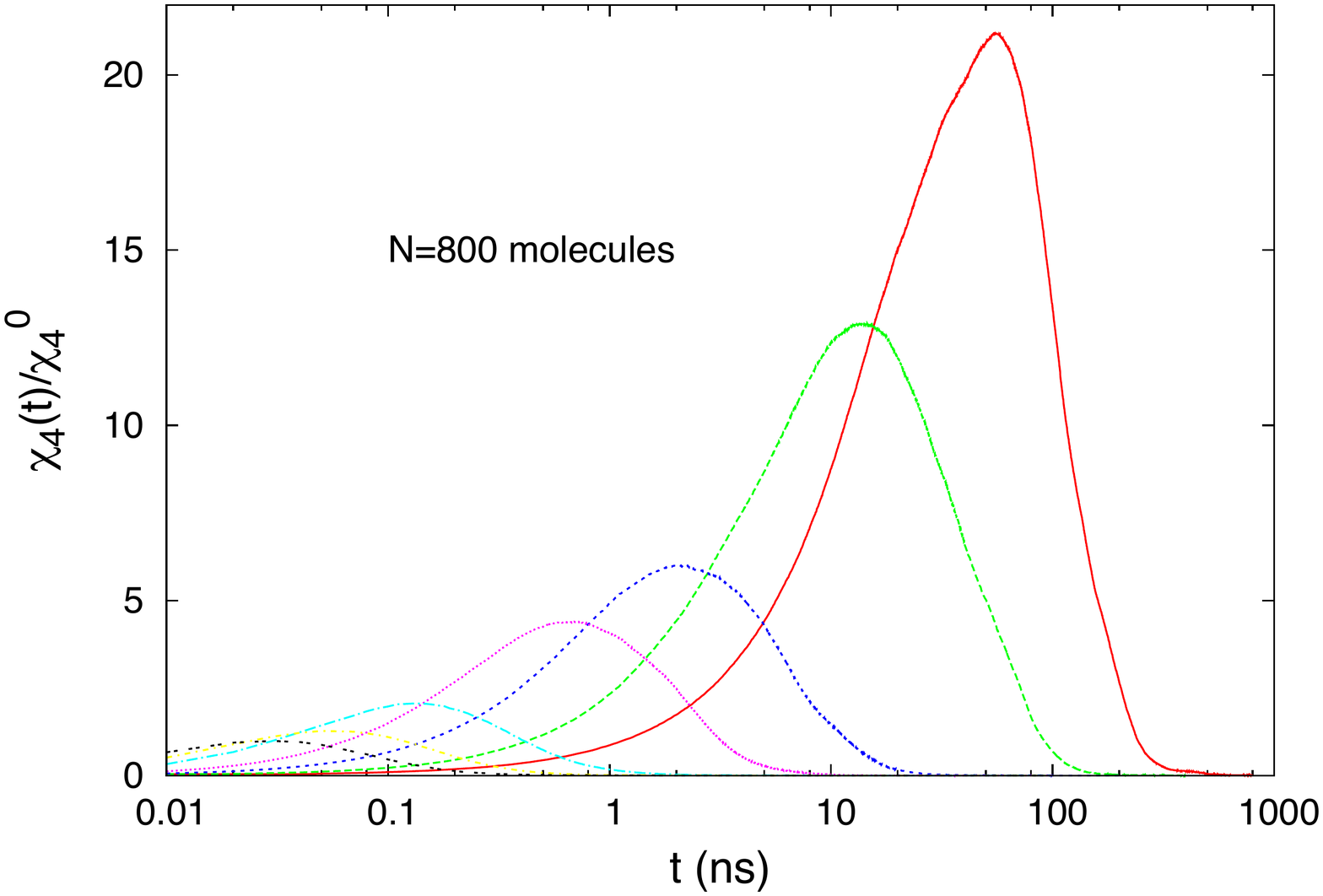}

{\em \footnotesize  FIG.4c. (color online) 4-point dynamic susceptibility $\chi_{4}$, for different temperatures. N=800 molecules. The temperature increases from right to left: T=115K (red curve), 120K (green curve), 130K (blue curve), 140K (purple curve), 160K (light blue curve), 180K (yellow curve) and 200K (black curve).\\}

In Figure 4 we show the evolution of the dynamic susceptibility $\chi_{4}(t)$  with the temperature and the system size.
We normalize the susceptibility with the corresponding maximum value $\chi_{0}$  in the largest system size (N=800) and temperature (T=200 K) investigated. Thus, the first points to the left in Figure 4a have a value equal to $1$.
Figures 4 show how the susceptibility increases when the temperature drops, which is a typical behavior in glass-formers, reflecting the increase of the cooperativity characteristic length. 
Above $T=150 K$ the susceptibility is the same for the various system sizes investigated (from $L/2=19.8$\AA $ $ to $10.9$\AA),  showing that the cooperativity length scale $\zeta$ is significantly smaller than our smaller system size for these temperatures (i.e. $\zeta < 10.9$\AA\ for $T \geq 150$ K).
Then as the temperature decreases the results corresponding to different system sizes split, the smaller box susceptibilities increasing less than those for the larger boxes.
This result suggests that as the temperature decreases, the increase of the cooperativity length scale progressively encounter the limits of larger boxes, leading to the split in the susceptibilities.  
This result thus confirms the picture of a cutoff of the cooperative length scale by the system size and shows the decrease of the susceptibility with the system size.

Together with that increase of $\chi_{4}(t)$ when the temperature drops, the maximum of the peak (Figures 4b and 4c) progressively shifts to larger times showing an increase of the characteristic time scales involved in the cooperative motions.
We also observe a decrease of the susceptibility  by a factor  $2$ when the half length of the simulation box decreases from $L/2=19.8$\AA $ $ to $10.9$\AA.
This result shows that the cooperative motions decrease significantly when the system size decreases, as expected in the picture of a cutoff introduced by the size of the system. 
Note that the small but continuous increase of the susceptibility that we observe in Figure 4b for a small box size (N=100 molecules) 
can be linked to the small non-Arrhenius behavior of $\tau_{\alpha}$ observed in Figure 1 with that box size. In that viewpoint, the small box simulations are consistent with the description of a different liquid with a smaller fragility.

\subsubsection{String-like motions:}

The cooperativity in supercooled liquids is partly represented by string-like cooperative motions\cite{strings,strings2,strings3} of the most mobile molecules. Molecules follow each other in strings, with a characteristic time delay  $\Delta t$ that increases when the temperature drops. The size of the strings has been associated to the size of the Cooperatively rearranging regions postulated by Adam et Gibbs theory in various works\cite{water,water2,adam3}.
To quantify the extent of string like motions in our liquid, we define the function $I_{string}(\Delta t)$ that measures the normalised number of molecules that follow other molecules in our simulation box, with a characteristic time $\Delta t$. 
The distinct part of the Van Hove correlation function
 \begin{equation}
\displaystyle{G_{d}(r,\Delta t)={1\over N} \sum_{i,j=1; i \neq j  }^{N} \delta(r-{\mid {\bf r}_{i}(\Delta t)-{\bf r}_{j}(0)\mid})} 
\end{equation}
 represents the probability to find a molecule at time $\Delta t$  a distance $r$ away from the location of another molecule at time $0$.
 When the temperature decreases a peak in this function develops around $r=0$ for a characteristic time $\Delta t=t^{*}$, showing that molecules are following each other on that characteristic time. 
We will here calculate $I_{string}$ by integrating that correlation function 
between $r=0$ \AA $ $ and $r=R_{c}=3.65$ \AA.
$I_{string}(\Delta t)$ is equal to zero for short times because $R_{c}$ is chosen smaller than the distance to the first neighbor, then $I_{string}(\Delta t)$ increases sharply  due to the molecules following each others, and eventually it decreases to a constant value for long time scales when the positions of the molecules are totally uncorrelated. 
 For this calculation we restrict ourselves to the 5 percent most mobile molecules in order to increase the proportion of string-like motions and the precision of our results.
We define the mobility of molecule $i$ at time $t$: $\mu_{i}(t,\Delta t)$, with a characteristic mobility time $\Delta t$ as:

\begin{equation}
\displaystyle{\mu_{i}(t,\Delta t)=  \mid{\bf r}_{i}(t+\Delta t)-{\bf r}_{i}(t))\mid }
\end{equation}
We then define $I_{string}(\Delta t)$ as:
\begin{equation}
\displaystyle{I_{string}(\Delta t)={1\over \rho}\int_{0}^{R_{c}} G_{d}(r,\Delta t)  4 \pi r^{2} dr}
\end{equation}

and define $I_{string}$ as the maximum of $I_{string}(\Delta t)$.

\includegraphics[scale=0.32]{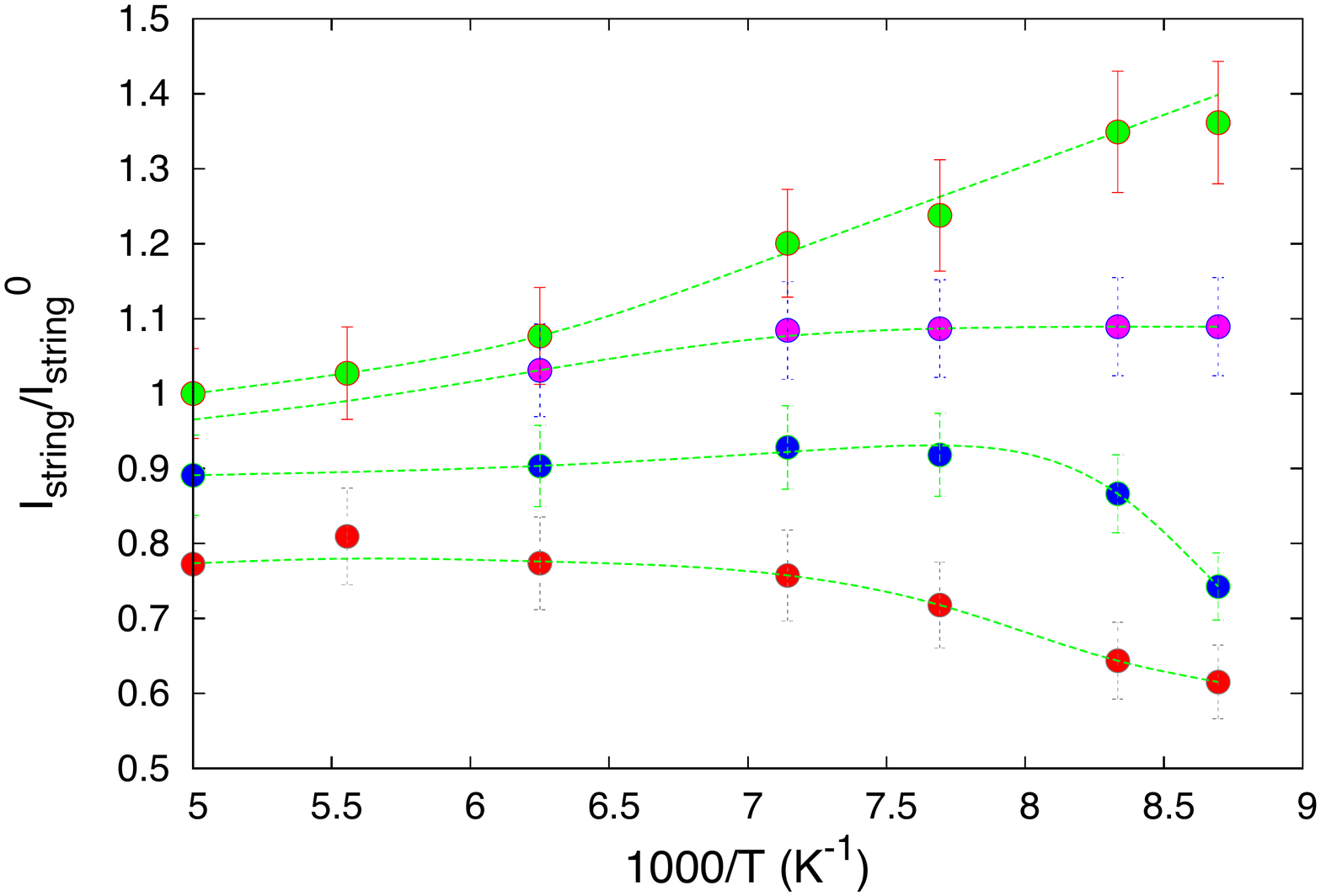}

{\em \footnotesize  FIG.5. (color online) Strength of the string-like motions for the $5$ percent most mobile molecules. $I_{string}^{0}$ is the value of $I_{string}$ for the largest system size (N=800) and temperature (T=200 K) investigated. From top to bottom, the sizes are: N=800 (green circles), N=400 (purple circles), N=200 (blue circles) and N=100 molecules (red circles). The lines are only guides to the eyes intended to clarify the Figure, and some caution must be taken for the interpretation of the Figure as other fits are possible within the error bars.\\}

Figure 5 shows the maximum values of $I_{string}(\Delta t)$, for various temperatures and system sizes.
As with the susceptibility, we normalize  $I_{string}$ with the corresponding value $I_{string}^{0}$ in the largest system size (N=800) and temperature (T=200 K) investigated.
The largest size plotted (N=800 molecules) displays the expected increase of string-like motions ($I_{string}$) as the temperature drops.
This result is in agreement with the increase of the susceptibility observed in Figure 4 at low temperatures and shows that the cooperativity of our system increases with $1/T$. 
Then as the system size  decreases, string-like motions cannot extend to distances larger than the box size and are thus inhibited. 
Note that due to their particular shape, strings of $n$ molecules are larger and thus more influenced by the system size than spherical aggregates of $n$ molecules.
We observe indeed in the Figure an important decrease of the string-like motions (larger than a factor $2$) as the system size decreases.
For a box containing $N=400$ molecules, the string motion strength $I_{string}$ stays roughly constant in the Figure.
For smaller boxes ($N=100$ to $200$ molecules), the string motion strength even decreases when the temperature drops.

\vskip 1cm

The number of molecules moving in strings has been associated in previous works to the system's configurational entropy\cite{water,water2,adam3} and to the size of the Cooperatively Rearranging Regions (CRR) postulated by the Adam-Gibbs theory\cite{adam}.
Within this picture a decrease of the strings size can be linked in the Adam-Gibbs model to the evolution of the relaxation times with the formula\cite{water,water2,adam3}:
\begin{equation}
{\displaystyle{\tau_{\alpha}=\tau_{0} e^{A/T S_{conf.}}= \tau_{0}  e^{A(n^{*}-1)/T}}}
 \end{equation}
 where $S_{conf}$ is the configurational entropy, $n^{*}$ is the average number of molecules inside a string, $T$ the temperature, while $A$ and $\tau_{0}$ are constants.
This formula leads to a decrease of the relaxation times when the strings' size decreases as observed in our simulations.

\vskip1cm
\subsubsection{Aggregation of the most or least mobile molecules:}

One of the main characteristic of cooperative motions called Dynamical heterogeneities (DH) is the observed structural heterogeneity in displacements. 
Molecules that move 'fast' (mobile molecules) segregate and 'slow' molecules also segregate. The terms 'fast' and 'slow' here do not refer to the instantaneous velocities but to the molecule's displacement (i.e. the mobility $\mu_{i}(t,\Delta t)$ ) on a characteristic time $\Delta t$ that increases when the temperature drops. Indeed instantaneous velocities are uniformly distributed, provided that the system is at equilibrium.

\includegraphics[scale=0.32]{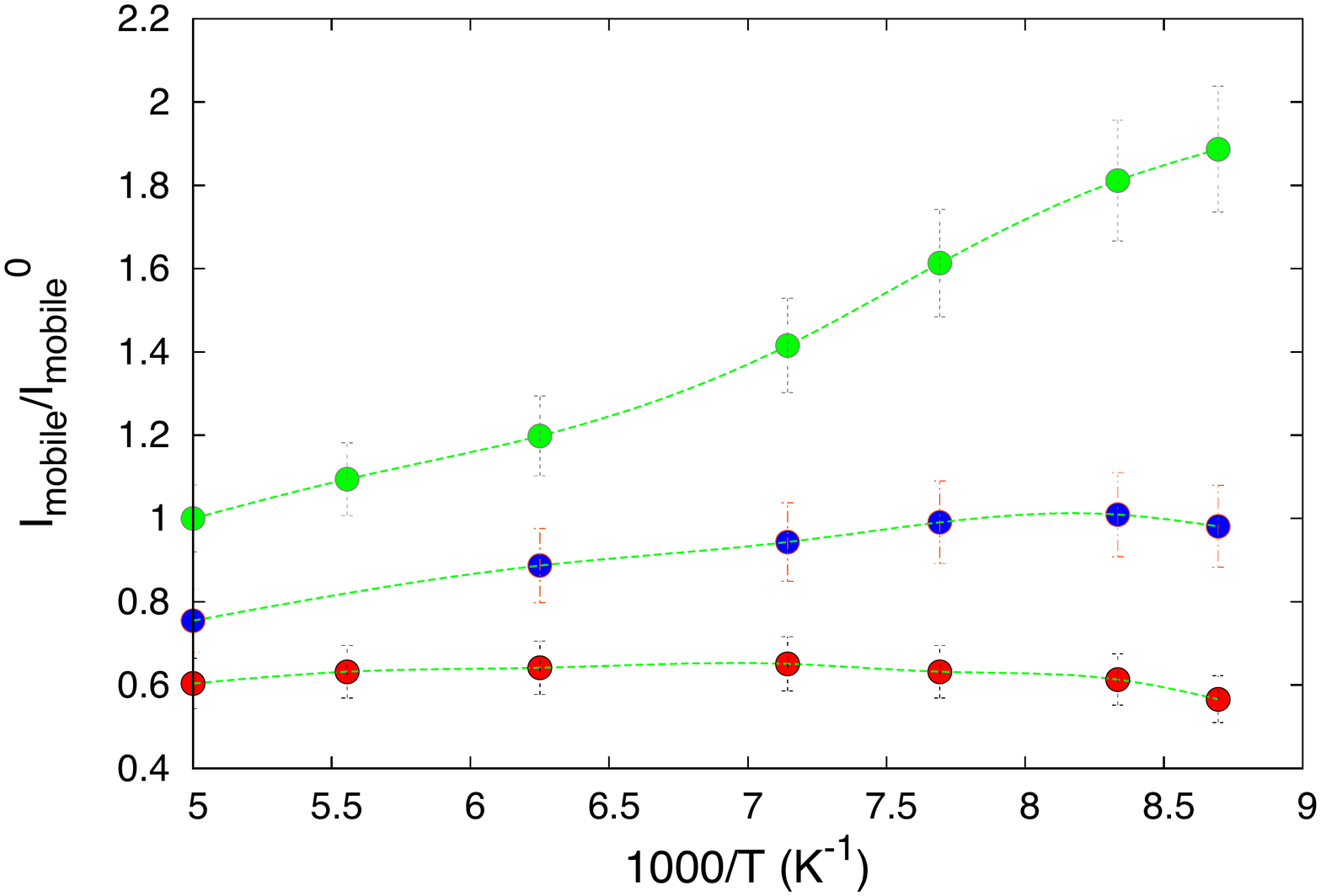}

{\em \footnotesize  FIG.6. (color online) Strength of the aggregation of the $5$ percent most mobile molecules. From top to bottom, the sizes are: N=800 (green circles), 200 (blue circles) and 100 molecules (red circles). The lines are guides to the eyes intended to clarify the Figure.\\}

Figure 6 and 7 show respectively the strength $I_{mobile}$ and $I_{slow}$ of the aggregation of the most or least mobile molecules in our liquid versus temperature for different system sizes. 
 Here $I_{mobile}$ and $I_{slow}$ are the maximum values of $I_{mobile}(\Delta t)$ and $I_{slow}(\Delta t)$
 that we obtain from the equations: 
\begin{equation}
{\displaystyle{I_{mobile}(\Delta t)=\int_{0}^{R_{c}} g_{mob-mob}(r) 4 \pi r^{2} dr/\int_{0}^{R_{c}} g(r) 4 \pi r^{2} dr   -1}}
\end{equation}

and similarly

\begin{equation}
{\displaystyle{I_{slow}(\Delta t)=\int_{0}^{R_{c}} g_{slow-slow}(r) 4 \pi r^{2} dr/\int_{0}^{R_{c}} g(r) 4 \pi r^{2} dr   -1}}
\end{equation}

The variable $\Delta t$ appears in these equations from the definition of the mobility $\mu_{i}(t,\Delta t)$ that we use to discriminate the 5 percent most or least mobile molecules. $g(r)$ is the radial distribution function between the centers of masses using the whole set of molecules, while $g_{mob-mob}(r)$ and $g_{slow-slow}(r)$ are the radial distributions between the most or least mobile molecules only. $R_{c}=7.05$\AA\ is a cutoff that we chose to be at the position of the first minimum of $g(r)$.

We observe in Figure 6 a behavior that is in between what we observed for string motions in Figure 5 and for the dynamic susceptibility in Figure 4.
For the largest box investigated (N=800 molecules) the aggregation strength $I_{mobile}$ increases roughly by a factor $2$ when the temperature drops, showing a large increase of cooperative motions. Then for smaller system sizes, $I_{mobile}$ decreases due to the cutoff generated by the limited box size on cooperative motions.
As for string motions we even observe a decrease of the dynamic heterogeneity  $I_{mobile}$ at low temperatures. 
Note that $I_{mobile}$ contains, but is not limited to, molecules participating to strings.

\includegraphics[scale=0.32]{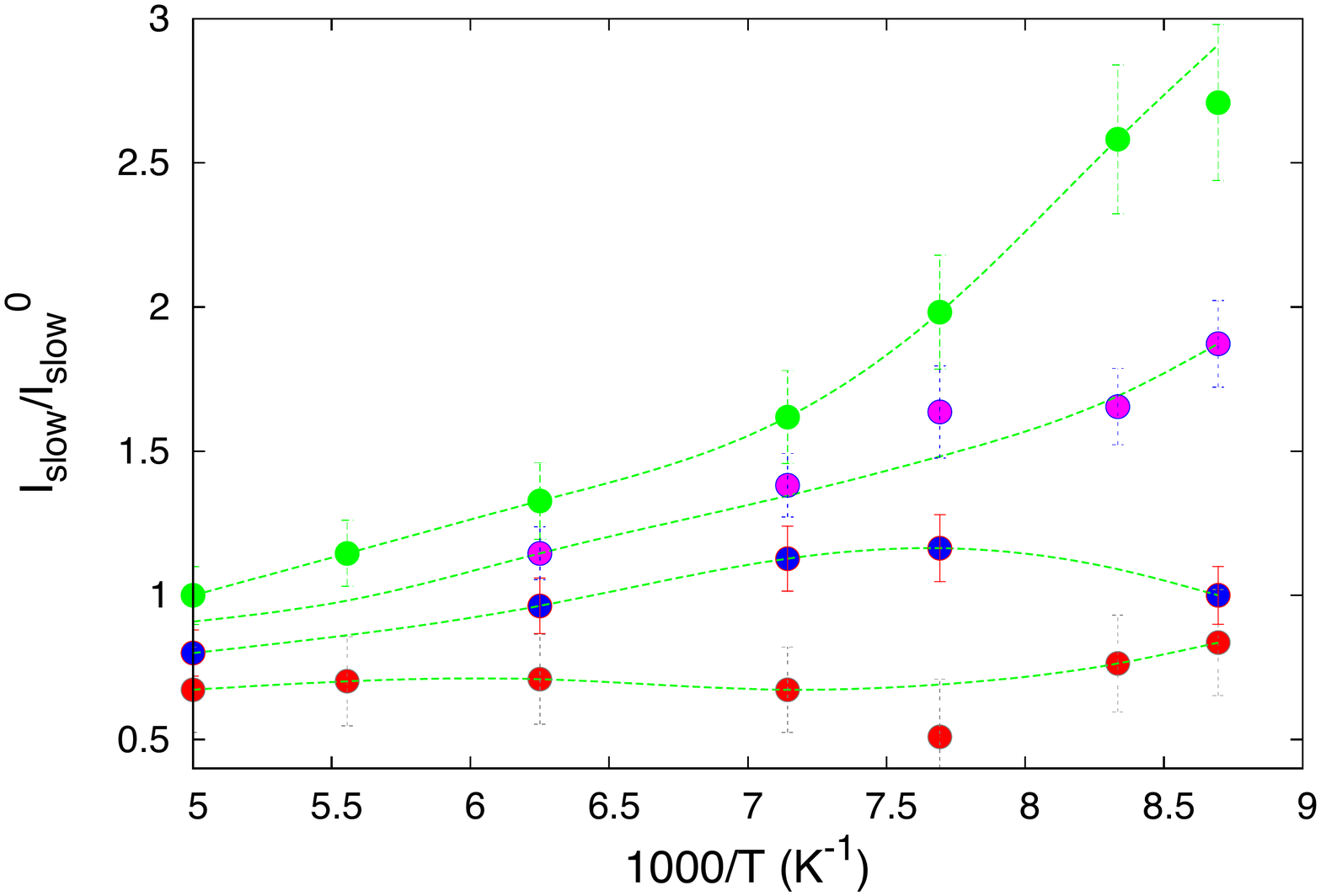}

{\em \footnotesize  FIG.7. (color online) Strength of the aggregation of the $5$ percent least mobile molecules. From top to bottom, the sizes are: N=800 (green circles), 400 (purple circles), 200 (blue circles) and 100 molecules (red circles). The lines are guides to the eyes intended to clarify the Figure.\\}

The strength $I_{slow}$ of the aggregation of the least mobile molecules in Figure 7 increases more rapidly (a factor 2.7 in the Figure) than $I_{mobile}$ when the temperature drops, for the largest box ( $I_{slow}^{N=800} > I_{mobile}^{N=800}$ ).
Then for smaller boxes, the aggregation of slow molecules decreases tending to a roughly constant behavior with temperature for the smallest box of the Figure. For the small boxes we observe an aggregation strength similar for the slow and mobile molecules ( $I_{slow}^{N=100} \approx I_{mobile}^{N=100}$ ).

\subsection{ Excitation concentration versus system size:\\}

The excitations, that are the elementary diffusive motions, are important quantities in the facilitation theory\cite{facilitation} but also can lead us to a better understanding of the physical processes that induce the diffusion.
We define the excitations as molecules moving  more than a threshold value $\Delta r$ during a time lapse $\Delta t$.  We choose $\Delta r$  large enough for the molecule to escape the cage and diffuse, and $\Delta t$ small enough to see elementary processes. Consequently we consider the molecule $i$ as excited at time $t$ if it fulfills the following condition: $\displaystyle{ \mid{\bf r}_{i}(t+\Delta t)-{\bf r}_{i}(t))\mid > \Delta r}$.
In what follows, we choose $\Delta t=10 ps$ and $\Delta r= 1.5$\AA.

\includegraphics[scale=0.32]{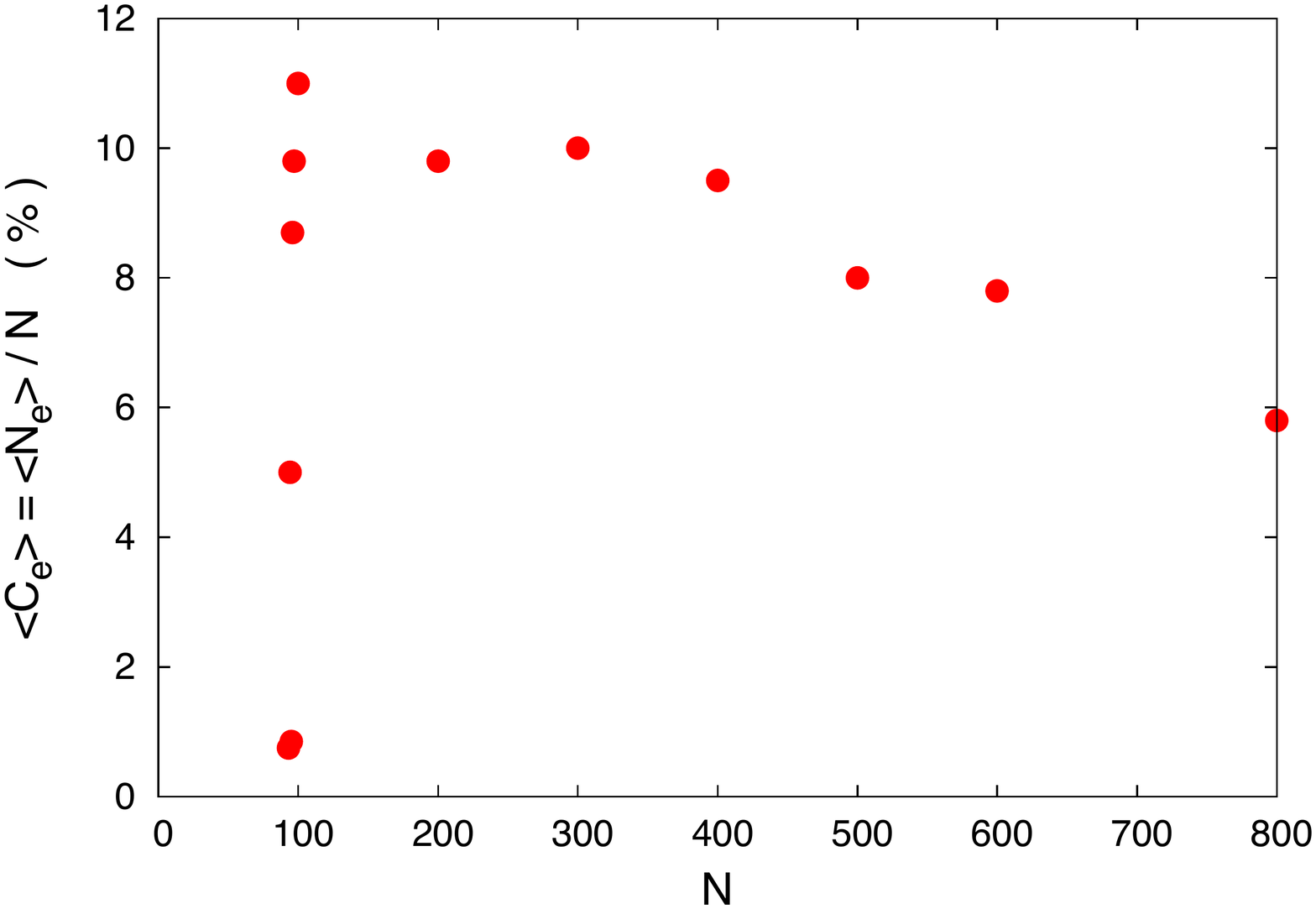}

{\em \footnotesize  FIG.8. (color online) Average excitation concentration in percent, versus the system size. The temperature is $T=120 K$.\\}

Figure 8 shows that the concentration of excitations increases slightly when the system size decreases, reaches a maximum for N=98 and then below N=97, the excitation concentration decreases abruptly.
The concentration of excitations thus follows the same trend than the diffusion coefficient when the size of the system decreases.
This result suggests that the increase of the diffusion when the system size decreases, arises from the increase of the concentration of excitations, and that the  abrupt decrease of the diffusion similarly arises from the decrease of the concentration of excitations. 
Figure 9 confirms that picture as the diffusion coefficient per excitation decreases when the system size increases, showing that the increase in the diffusion coefficient arises from the increase in the concentration of excitations and not from an increase of the diffusion per excitation. 
We explain that continuous decrease of the diffusion per excitation for $N>98$ as arising from the preferential cutoff of large cooperative string-like motions.
Since the larger cooperative motions are also the fastest (i.e. correspond to the most mobile molecules)\cite{strings}, the preferential cutoff of these motions induces a decrease in the average mobility per excitation.
  
\includegraphics[scale=0.32]{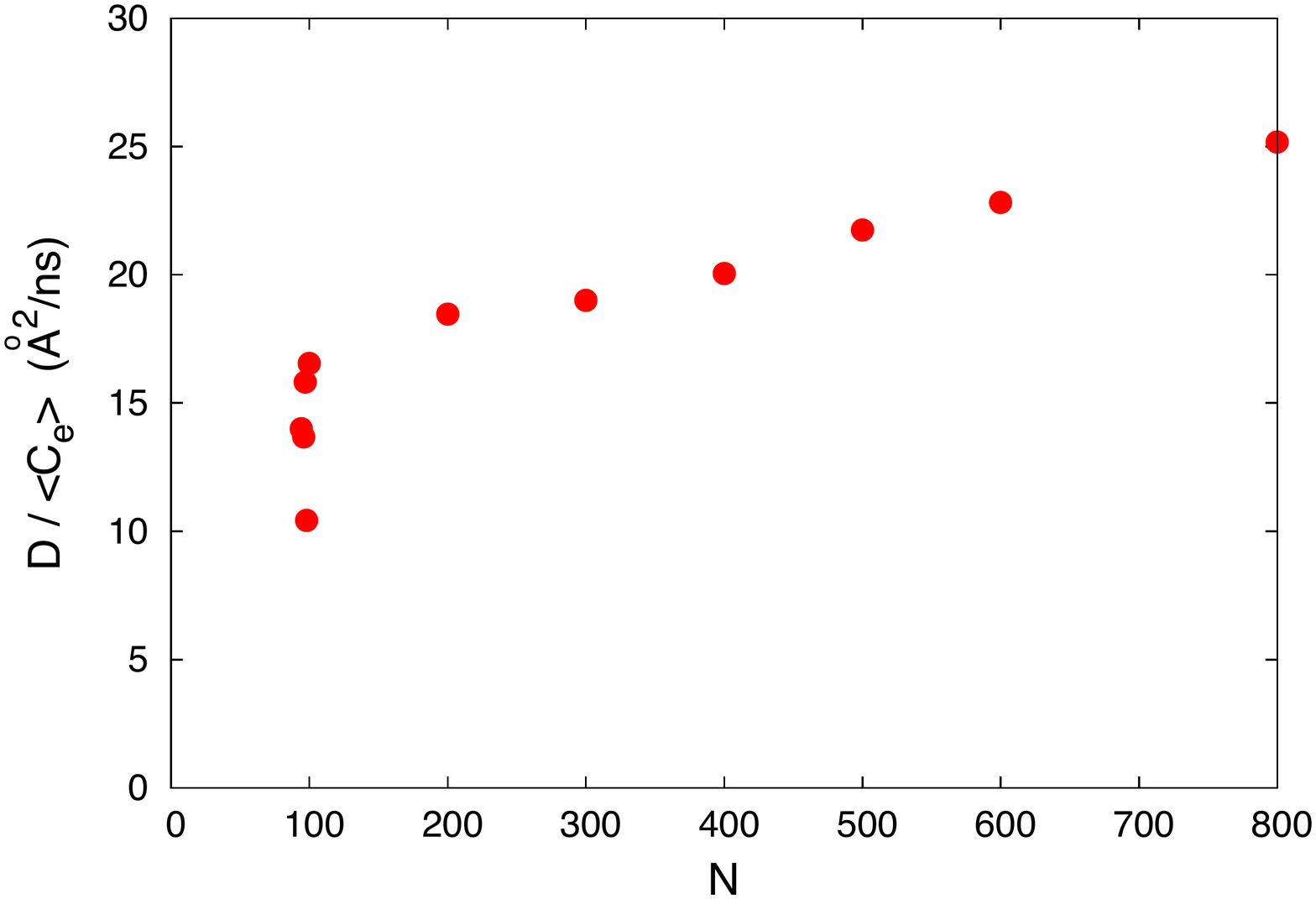}

{\em \footnotesize  FIG.9. (color online)  Diffusion coefficient divided by the average excitation concentration versus the system size.
$<r^{2}(t)>=<C_{e}><r^{2}(t)_{e}>+(1-<C_{e}>)<r^{2}(t)_{non.e}>$, leading for large $t$ values to $<r^{2}(t)>\approx<C_{e}><r^{2}(t)_{e}>$. 
Consequently, since $D=lim_{t -> \infty} <r^{2}(t)>/6 $, $D/<C_{e}>$ quantifies the average contribution to the diffusion induced by an excitation. 
When the system size decreases this average contribution decreases, a result that we interpret as coming from the cutoff of long cooperative chains that correspond to the larger diffusive motions\cite{dh4}. The temperature is $T=120 K$.\\}

We will now search an explanation for the evolution of the excitation concentration and diffusion coefficient in our system.
When the system size decreases the relative fluctuations in the number of excitations increases, leading for small sizes intermittently to a very small number of excitations or even no excitation inside the system. When it happens it induces an abrupt decrease of the diffusion coefficient. Then as the motion is stopped this behavior persists and the concentration of excitations averaged on time continue to decrease.
That behavior is in qualitative agreement with the facilitation theory. 

\includegraphics[scale=0.32]{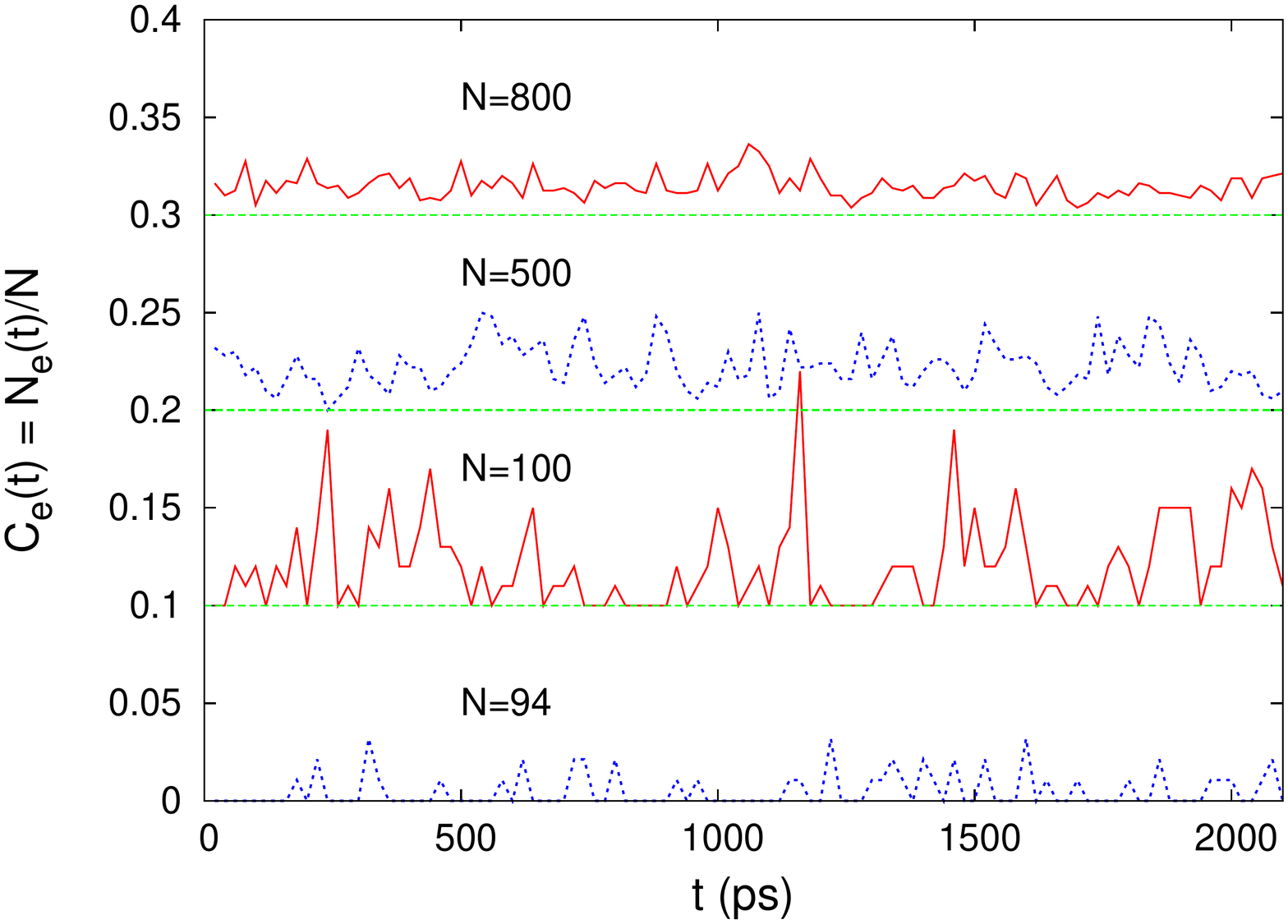}

{\em \footnotesize  FIG.10. (color online) Time dependence of the excitation concentration for different system sizes. To separate the different sizes, we have added the value 0.1 between each graph concentrations. Thus each graph concentration is in reality comprised in between 0. and 0.1 and the green dashed lines correspond to the zero of the next graph. The temperature is $T=120 K$. \\}

To test that explanation we display in Figure 10 the time dependence of the excitation concentration. The Figure shows an increase of the fluctuations when the size decreases, with large fluctuations for N=100 and an intermittent behavior for the smaller size ($N=94$). These results thus confirm the picture of the evolution of the diffusion arising from the fluctuations of the excitations.

The slight increase of the concentration of excitations and as a result of the diffusion, observed for larger decreasing sizes, is more difficult to explain.
When the size decreases it results in a cutoff of cooperative motions. However as discussed above, the larger cooperative string like motions, that will be preferentially cut off due to their size, correspond to the fastest motions, leading to the decrease of the average diffusion per excitation with decreasing $N$ that we observe in Figure 9.
We note that this decrease could also explain the decrease of the diffusion observed in other systems.  
We tentatively explain the increase of the excitation concentration in our system as due to an increase of smaller size cooperative motions as the large strings are cut off.

\section{Conclusion}

In this work we have studied the finite size dependence of the dynamical and structural properties of a very simple glass-former made of linear dumbbell Lennard-Jones molecules. Our aim was to investigate the relation between the cooperativity length scale, that the system size modifies, and the dynamics of the liquid.
Functions measuring cooperative motions, such as the dynamic susceptibility, the strength of the aggregation of the most mobile or least mobile molecules, and the strength of string motions all were found to display an increase of their maximum values as the temperature decreases. 
All these functions also display a decrease of their maxima when the size of the system decreases.
We concluded that when the system size decreases, cooperative motions cannot extend to larger distances than the box size, leading to the observed decrease of the cooperativity. These results agree well with previous works using finite size to investigate the presence of cooperative length scales in various glass-formers.
Comparing that decrease of the cooperativity with the evolution of the relaxation time of the material, we found that the  relaxation time, like the cooperativity, decreases when the size of the box  decreases. 
However we note that this result is in opposition with what was found previously in different glass-formers like silica, and Kob Andersen mixtures of Lennard-Jones atoms\cite{size0,size1,size2,size3,size4}.
Finally we found that the finite size dependence of the transport properties is non-monotonic suggesting the presence of two different competing physical mechanisms with different length scales, a result that may explain the opposite tendency found in previous works for the relaxation times evolution with the system size.

 \vskip 1cm
 
 {\bf APPENDIX: Simulation details\\}

Simulations of small systems are usually more affected by thermostats or  simulation algorithms than simulations of larger systems.
Thus we have tested  the effect of the variation of the simulation time step and of the Berendsen thermostat parameter on our main results.
We also have tested the effect of including the box replica to improve the statistical accuracy in a few particular cases and show these data here. We study different system sizes, but we focus our attention on the small system size $N=100$ that corresponds approximately to the maximum of the diffusion in our simulations. We did not find significant variations of the diffusion coefficient and cooperative motions in that study.

\vskip 0.5cm
{\bf Time step variation\\}

\includegraphics[scale=0.33]{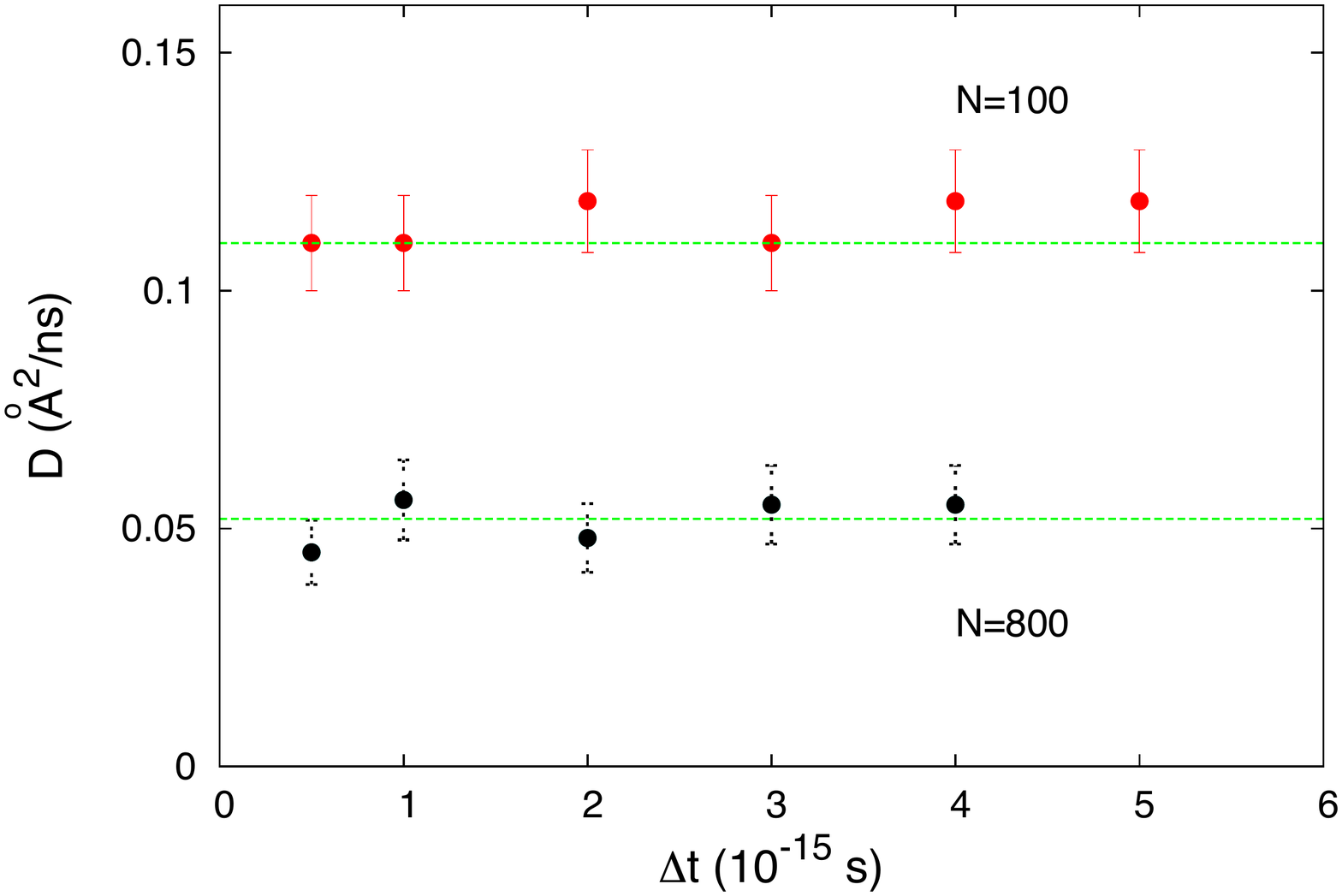}

{\em \footnotesize  FIG.11(a). (color online)  Diffusion coefficient versus the simulation time step $\Delta t$. The precision of the calculations and the CPU time both increase for small time steps leading usually to search for a compromise, but we have used in our work a relatively small time step ($\Delta t=10^{-15}s$). The lines are guides to the eyes. The temperature is T=120K.\\}

Figure 11a shows the diffusion coefficient obtained using different time steps $\Delta t$ ranging from $5$ $10^{-16}s$ to $5$ $10^{-15}s$ in the simulations.
For larger time steps the simulations are usually not possible with predictor-corrector algorithms.
For large time steps  we expect the results to be affected due to the decrease in the precision of the algorithm.
However these results show that there is no significant effect on the diffusion coefficient for $\Delta t \leq 5$ $10^{-15}s$.

\includegraphics[scale=0.33]{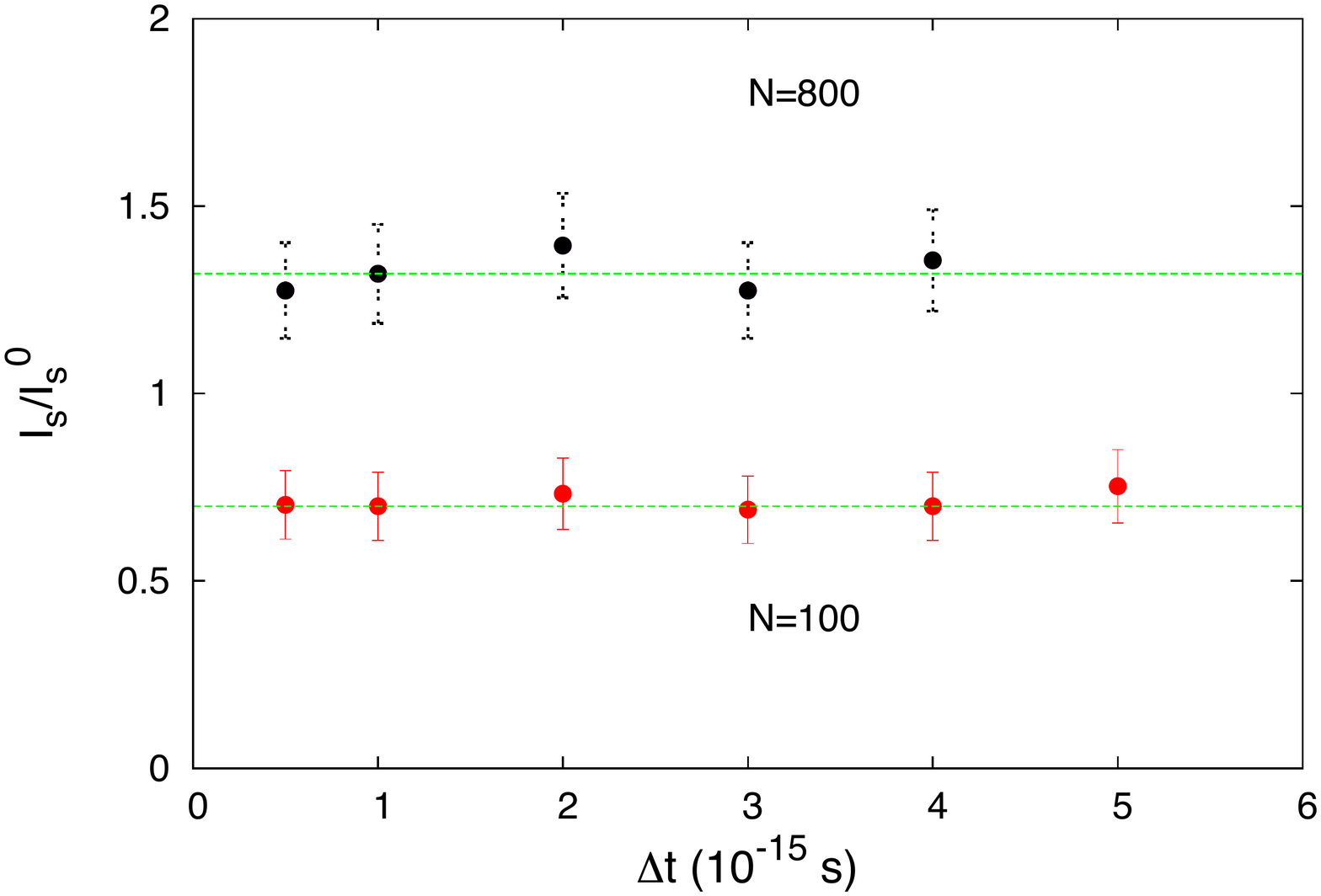}

{\em \footnotesize  FIG.11(b). (color online)  Strength of string like motions versus the simulation time step $\Delta t$.  The temperature is T=120K. The lines are guides to the eyes.\\}

The strength of string like motions in Figure 1b also show no significant variation for the different time steps displayed.

\vskip 0.5cm
{\bf Thermostat parameter\\}

Figure 12a shows the diffusion coefficient obtained from simulations using different strength parameters ($1/ \tau_{Berendsen}$) for the thermostat.
A decrease in  $1/ \tau_{Berendsen}$ corresponds to a decrease in the strength of the thermostat, and  $1/ \tau_{Berendsen}=0$ corresponds to the microcanonical limit (i.e. no thermostat). The values in the Figure correspond to a parameter $\tau_{Berendsen}$ ranging from $10^{-15} s$ to $8$ $10^{-13} s$ (i.e. $1/ \tau_{Berendsen}$ ranging from $1.2$ $10^{12} s^{-1}$ to $10^{15} s^{-1}$. We have used a strength parameter $1/\tau_{Berendsen}=10^{13} s^{-1}$ in our work.
We expect spurious effects to appear for large values of the parameter $1/ \tau_{Berendsen}$ as the thermostat will decrease the thermal fluctuations in the system.
This effect is expected to be larger for small system sizes, due to the decrease in the number of degrees of liberty for small systems that will lead to an increase of the thermostat effect on the molecular motions.
We see in Figure 12a that a thermostat effect appears for N=100 when the strength parameter of the thermostat is larger than $1/ \tau_{Berendsen} \approx 10^{14} s^{-1}$.
Finally, Figure 12b shows that  the strength of the string like motions stays constant when we vary the thermostat parameter within the strength range of the Figure.

\includegraphics[scale=0.33]{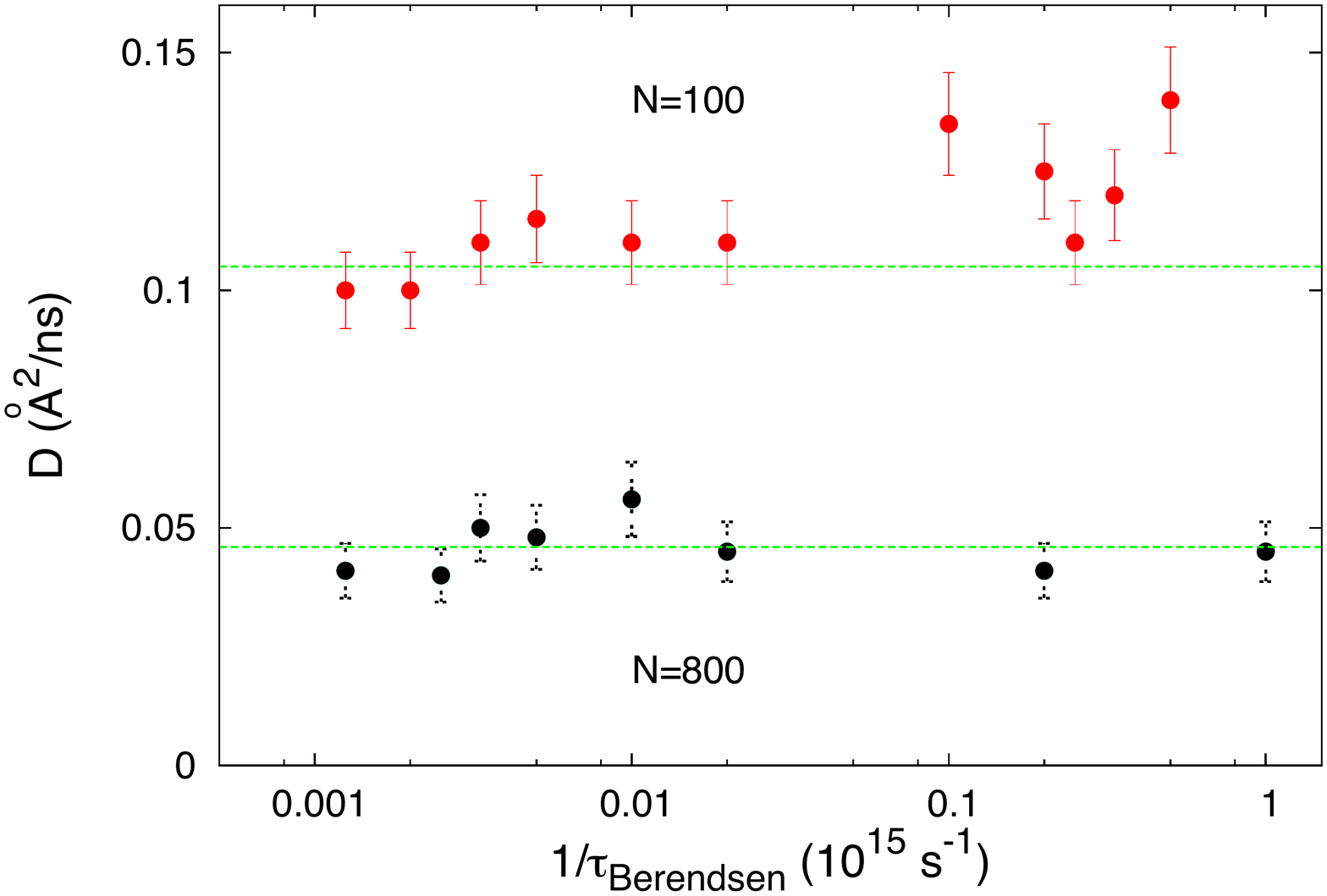}

{\em \footnotesize  FIG.12(a). (color online) Diffusion coefficient versus the inverse of the thermostat constant $\tau_{Berendsen}$.
A decrease in  $1/ \tau_{Berendsen}$ corresponds to a decrease in the strength of the thermostat and 
$1/ \tau_{Berendsen}=0$ corresponds to the microcanonical limit (i.e. no thermostat).  The temperature is T=120K.  \\}

\includegraphics[scale=0.33]{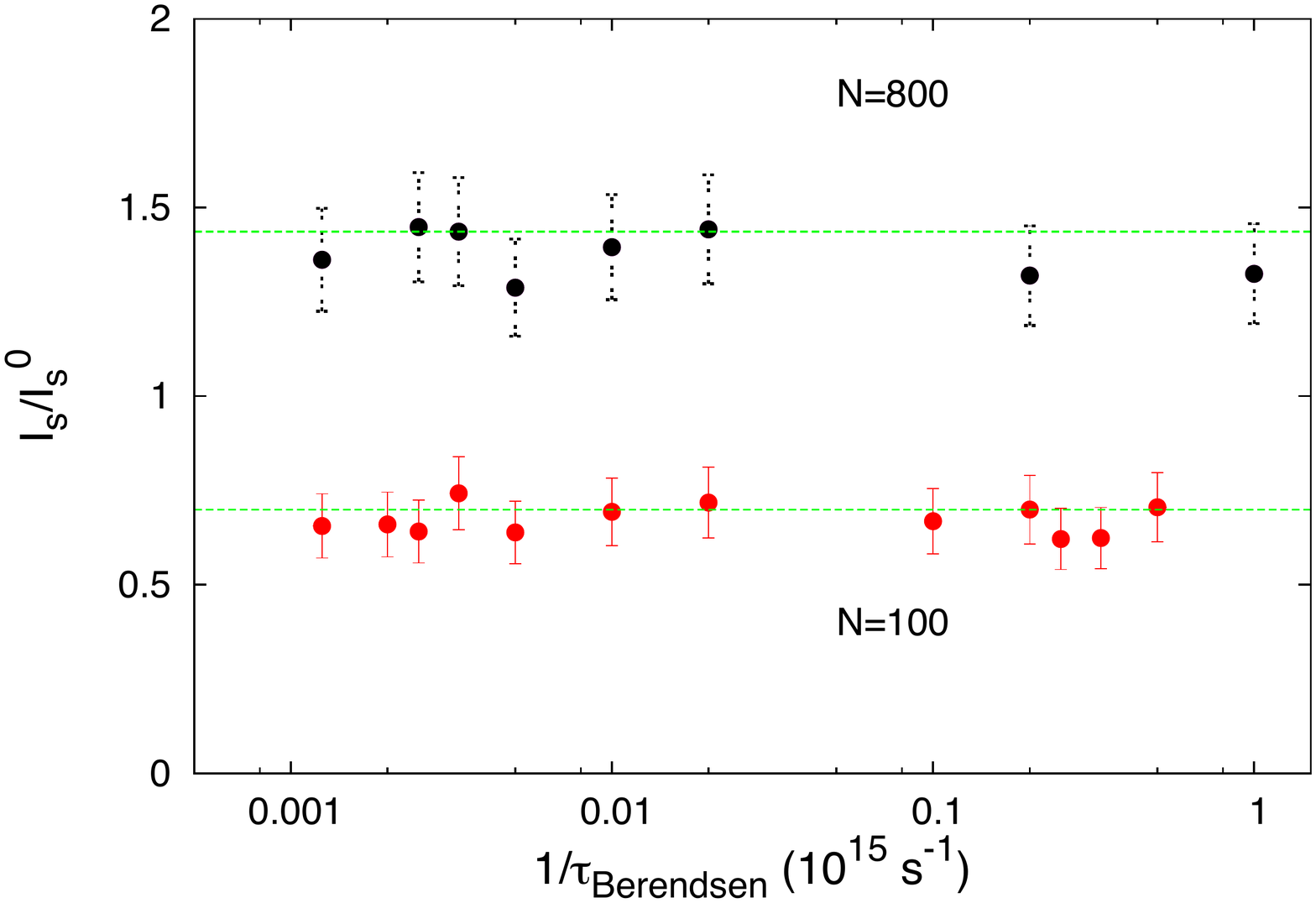}

{\em \footnotesize  FIG.12(b). (color online) Strength of string like motions versus the inverse of the thermostat constant $\tau_{Berendsen}$.
The temperature is T=120K.\\}

\vskip 0.5cm
{\bf Simulation box replica\\}

In this section we show the effect of adding the first replica of the simulation box to improve the statistics in the calculation of the correlation functions.  
While the diffusion coefficient and the susceptibility cannot be improved with that method as they use correlations between the successive positions of a molecule,
the functions based on radial distribution functions calculations or on distinct Van Hove correlation functions can be improved slightly with that method. 
We show in Figure 13 the radial distribution function (RDF) with or without the contribution of the $26$ first replica of the box.
The calculations are limited here to $r<L/2$ when only the simulation box is used, and to $r<L$ when the replica are used.
Consequently the calculations do not include the correlation between the molecule and its replica.
The Figure shows that within that range of $r$ there are no spurious effects and the RDF thus could be extended to increase the precision by using the replica.
Results shown in the manuscript use these replica to increase the precision, using the limit $r<0.75 L$.
As expected the RDF are exactly the same for $r<L/2$.

\includegraphics[scale=0.33]{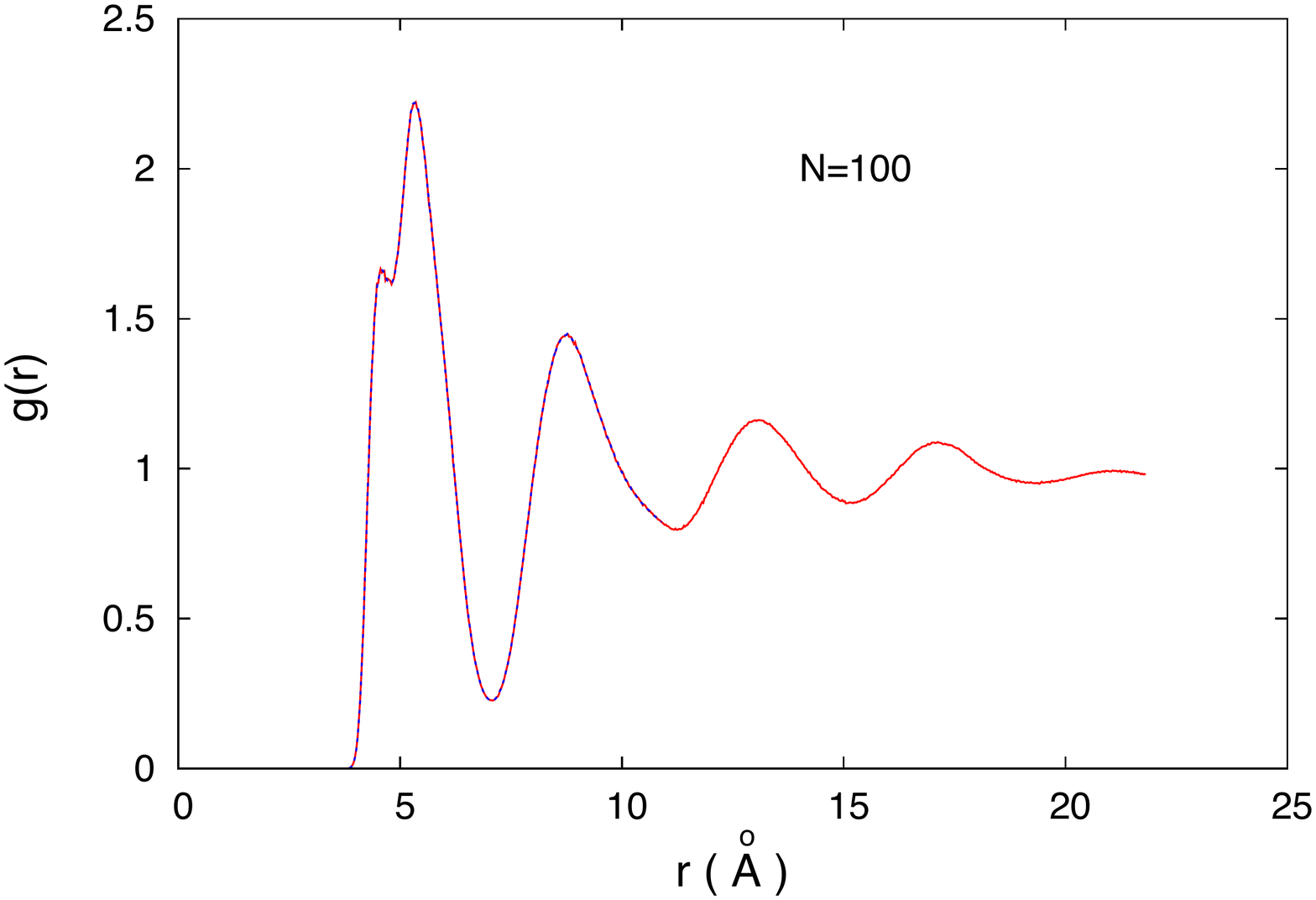}

{\em \footnotesize  FIG.13. (color online) Radial distribution function between the centre of masses, using molecules from the simulation box (blue line) or molecules from the simulation box and the first $26$ box replica (red line). The calculations are limited here to $r<L/2$ when only the simulation box is used, and to $r<L$ when the replica are used. $N=100$ molecules and $T=120 K$.
The temperature is T=120K.\\}

The effect of using the replica on the distinct Van Hove correlation function, using the $5$ percent most mobile molecules, is shown in Figure 14.
We see the appearance of a peak around $r=0$\AA$ $ showing that molecules are following each other on the characteristic time $t^{*}$ used in the calculation.
The presence of that peak is a characteristic of string like motions and used in the manuscript to calculate the functions $I_{s}$ that measure the strength of the string  motions. The Figure shows quite small differences between the calculation using the replica method and the calculation using only the simulation box.
These small differences arise from the contribution of the molecules of the simulation box that are located at $r>L/2$ (i.e. that are in the corners of the box). 
These contributions increase slightly the accuracy of the calculation.

\includegraphics[scale=0.33]{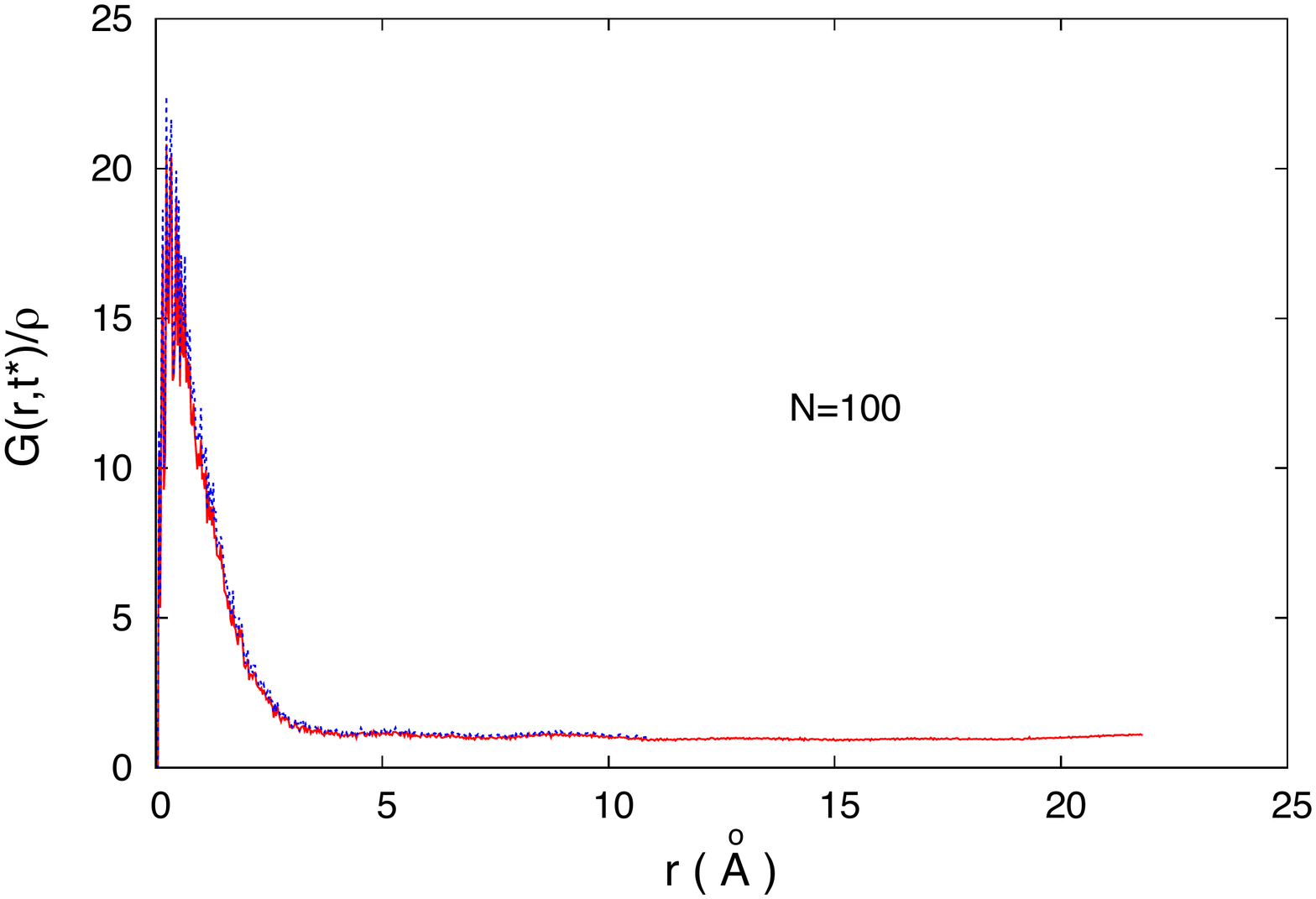}

{\em \footnotesize  FIG.14. (color online) Distinct part of the Van Hove correlation function, using the $5$ percent most mobile molecules from the simulation box (blue line) or molecules from the simulation box and the first $26$ box replica (red line). As in Figure 13, the calculations are limited  to $r<L/2$ when only the simulation box is used, and to $r<L$ when the replica are used. $N=100$ molecules and $T=120 K$. The peak around $r=0$\AA$ $ shows that molecules are following each other on the characteristic time $t^{*}$ that corresponds to the maximum of the Non Gaussian parameter. Functions $I_{s}$ correspond to the integration of that peak. $\rho$ is the liquid density.
The temperature is T=120K.\\}

\end{document}